\documentclass[12pt,preprint]{emulateapj}
\usepackage{natbib}
\usepackage{color}
\usepackage{hyperref} 
\hypersetup{colorlinks=true,citecolor=blue}
\usepackage[T1]{fontenc}
\usepackage{aecompl}
\def \bea {\begin{eqnarray}}
\def \ena {\end{eqnarray}}                  
\def \bee {\begin{equation}}
\def \ene {\end{equation}}
\def    \simlt  {\lower.5ex\hbox{$\; \buildrel < \over \sim \;$}}
\def    \simgt  {\lower.5ex\hbox{$\; \buildrel > \over \sim \;$}}

\newcommand     \mum    {\,\mu{\rm m}}  

\def	\cm		{\,{\rm {cm}}}

\def	\km		{\,{\rm {km}}}

\def	\erg		{\,{\rm {ergs}}}

\def    \exp 		{\,{\rm {exp}}}
\def	\g		{\,{\rm g}}

\def	\pc		{\,{\rm {pc}}}

\def	\s		{\,{\rm s}}

\def    \ln  		{\,{\rm {ln}}}

\def	\mag		{\,\rm mag}

\def	\H		{\rm H}

\def	\V		{\rm V}

\def	\gas		{\rm {gas}}

\def    \ext    	{\rm {ext}}
\def    \pol    	{\rm {pol}}

\def	\gra	{\rm {gra}}
\def	\sil	  {\rm {sil}}
\def	\carb	{\rm {carb}}

\def	\obs		{\rm {obs}}

\def	\ali		{\rm {ali}}

\begin{document}
\title{Properties and alignment of interstellar dust grains toward Type Ia Supernovae with anomalous polarization curves}
\author{Thiem Hoang \altaffilmark{1,}\altaffilmark{2,}\altaffilmark{3}
}
\altaffiltext{1}
{Korea Astronomy and Space Science Institute, Daejeon, 305-348, Korea; \href{mailto:thiemhoang@kasi.re.kr}{thiemhoang@kasi.re.kr} }
\altaffiltext{2}
{Canadian Institute for Theoretical Astrophysics, University of Toronto, 60 St. George Street, Toronto, ON M5S 3H8, Canada}
\altaffiltext{3}
{Institute of Theoretical Physics, Goethe  Universit$\ddot{\rm a}$t Frankfurt, D-60438 Frankfurt am Main, Germany}

\begin{abstract}

Recent photometric and polarimetric observations of type Ia supernovae (SNe Ia) show unusually low total-to-selective extinction ratio ($R_{V}<2$) and wavelength of maximum polarization ($\lambda_{\max}<0.4\mum$) for several SNe Ia, which indicates peculiar properties of interstellar (IS) dust in the SN hosted galaxies and/or the presence of circumstellar (CS) dust. In this paper, we use inversion technique to infer best-fit grain size distribution and alignment function of interstellar grains along the lines of sight toward four SNe Ia with anomalous extinction and polarization data (SNe 1986G, 2006X, 2008fp, and 2014J). We find that to reproduce low values of $R_{V}$, a significant enhancement in the mass of small grains of radius $a< 0.1\mum$ is required. For SN 2014J, a simultaneous fit to its observed extinction and polarization is unsuccessful if the entire data are attributed to IS dust (model 1), but a good fit is obtained when accounting for the contribution of CS dust (model 2). For SN 2008fp, our best-fit results for model 1 show that, to reproduce an extreme value of $\lambda_{\max}\sim 0.15\mum$, small silicate grains must be aligned as efficiently as big grains. For this case, we suggest that strong radiation from the SN can induce efficient alignment of small grains in a nearby intervening molecular cloud via radiative torque mechanism. The resulting time dependence polarization from this RAT alignment model can be tested by observing at ultraviolet wavelengths. 

\end{abstract}

\section{Introduction}\label{sec:intro}
Type Ia supernovae (SNe Ia) have been used as standard candles to measure the expansion of the universe due to their stable intrinsic luminosity (\citealt{1998AJ....116.1009R}). Recently, a new avenue appears in using SNe Ia to study properties of dust in the interstellar medium (ISM) of distant galaxies (\citealt{2008A&A...487...19N}; \citealt{Foley:2014br}; \citealt{2014arXiv1408.2381B}). {This advancement is based on the strong dependence of peak luminosity on the reddening of SNe Ia by dust extinction (see \citealt{2013ApJ...779...38P})}.

Photometric observations of SNe Ia demonstrate peculiar dust properties with unprecedented low visual-to-selective extinction ratio $R_{V}=A_{V}/E_{B-V}$ where $A_{V}$ is the optical extinction and $E_{B-V}$ is the reddening (i.e., color excess). For instance, using data from 80 SNe Ia with considerable reddening (i.e., $E_{B-V} < 0.87\mag$), \cite{2008A&A...487...19N} report an unusually low value of $R_{V}\sim 1.75$ (see also \citealt{2013ApJ...779...38P}), much lower than the typical value $R_{V}\sim 3.1$ for interstellar grains in the Milky Way. Recently,  using data from UV to near-IR ($\lambda\sim 0.2-2\mum$) from Hubble and Swift satellites, \cite{2015MNRAS.453.3300A} present a diversity in extinction curves, with $R_{V}$ ranging from 1.4 to 3 (see also \cite{2012ApJ...749..126W}). Unprecedented low values of $R_{V}$ toward numerous SNe Ia are attributed to the scattering by circumstellar (CS) dust (see, e.g. \citealt{2008ApJ...686L.103G}). Numerous authors, however, suggest that the low values of $R_{V}$ are due to the enhancement in the relative abundance of small grains of interstellar (IS) dust in the host galaxy (see \citealt{2013ApJ...779...38P}. Thus, it is interesting to infer size distribution of interstellar grains that reproduce such low values of $R_{V}$.

Polarimetric studies of SNe Ia have opened a new window into probing dust properties and grain alignment in the galaxies because the intrinsic polarization of SNe light is negligible (see \citealt{2008ARA&A..46..433W} for a review). Four SNe Ia with high quality of polarization data (SNe 1986G, 2006X, 2008fp, and 2014J) exhibit anomalous polarization curves that rise toward UV wavelengths (\citealt{Kawabata:2014gy}; \citealt{Patat:2015bb}). Fitting the observational data with the typical ISM polarization law--Serkowski law--yields an anomalous value of peak wavelength ($\lambda_{\max}<0.2\mum$) for SNe 2008fp and 2014J, which is much lower than the typical value $\lambda_{\max}\sim 0.55\mum$ in the Galaxy. For comparison, available data for Galactic polarization show the lowest peak wavelengths $\lambda_{\max}=0.33\mum$ and $0.35\mum$ for Cyg OB2 No. 10 and 12 (\citealt{Whittet:1992p6073}), and $\lambda_{\max}<0.35\mum$ for HD 193682 \citep{1996AJ....112.2726A}.

The question of which origin (IS dust or CS dust) for these anomalous values of $R_{V}$ and $\lambda_{\max}$ remains unclear, and addressing this question has important implication for better understanding the progenitors of SNe Ia. If IS dust is important as suggested in previous studies (\citealt{2013ApJ...779...38P}; \citealt{Patat:2015bb}), the anomalous values could reflect peculiar properties of dust and alignment of grains in intervening clouds along the lines of sight to SNe or perhaps in the average IS dust of the hosted galaxies (\citealt{2013ApJ...779...38P}; \citealt{Patat:2015bb}). In this paper, we will use inversion technique (Section \ref{sec:method}) to infer best-fit grain size distribution and alignment function, in order to understand dust properties and alignment underlying the anomalous features. 

The structure of the paper is as follows. In Section \ref{sec:obs} we present observational data for the selected SNe Ia compiled from the literature. Section \ref{sec:theory} is devoted to present theoretical models of extinction and polarization and observational constraints. In Section \ref{sec:method} we briefly describe our inversion technique and obtained results. Extended discussion on dust properties and grain alignment mechanisms responsible for the anomalous extinction and polarization are presented in Section \ref{sec:dis}. A short summary is given in Section \ref{sec:summ}.

\section{Observed Extinction and Polarization Data}\label{sec:obs}

\subsection{Extinction and Polarization of Starlight in the Milky Way}

Interstellar dust causes extinction of starlight, which can be described by \cite{1989ApJ...345..245C} (CCM) extinction law with $R_{V}\sim 3.1$. In addition, the alignment of non-spherical grains with interstellar magnetic fields produces differential extinction of starlight, resulting in starlight polarization, and thermal emission from aligned dust grains become linearly polarized (see \citealt{Andersson:2015bq}; Lazarian, Andersson, \& Hoang 2015 (LAH15) for latest reviews).

The wavelength-dependence polarization (hereafter polarization curve) induced by IS dust is approximately described by the Serkowski law \citep{Serkowski:1975p6681}:
\bea
P_{\rm is}(\lambda)=P_{\max}\exp\left[-K\ln^{2}\left(\frac{\lambda_{\max}}{\lambda}\right)\right],\label{eq:serkowski}
\ena
where $P_{\max}$ is the maximum polarization at wavelength $\lambda_{\max}$, and $K$ is a parameter {related to the width of the Serkowski law} (\citealt{1980ApJ...235..905W}). For the Galaxy, $\lambda_{\max}\sim 0.55\mum$, although some lines of sight (LOS) exhibit lower $\lambda_{\max}$. But to date, the low values of $\lambda_{\max}<0.4\mum$ have not been detected (see \citealt{2007ApJ...665..369A}), {except a few lines of sight toward SNe Ia mentioned here}.

\subsection{SNe 1986G and 2006X}

SN 1986G exploded in a dust lane in the host galaxy NGC 5128 (Cen A), which was found to have at least a dozen of distinct clouds along the line of sight (\citealt{1989A&A...215...21D}; \citealt{1992A&A...259...63C}). Optical and near-IR photometric observations show that the extinction of SN 1986G can be described by the CCM law with $R_{V}=2.57$ (see \citealt{2013ApJ...779...38P}). Polarimetric measurements were first reported in \cite{Hough:1987p6065}, in which the polarization curve is well fitted by the Serkowski law with $\lambda_{\max}=0.43\mum$. {\cite{Hough:1987p6065} suggested that the average size of dust grains in Cen A is perhaps smaller than in the Milky Way.} It was also suggested that interstellar grains in the dust lane are aligned with the local magnetic field but with the alignment efficiency lower than that in the Milky Way. 

SN 2006X exploded within or behind the disk of NGC 4321 normal galaxy in the Virgo cluster. Photometric observations in optical and near-infrared by \cite{2008ApJ...675..626W} reveal an unusually low value of $R_{V}\approx 1.5$, and spectropolarimetric data \citep{Patat:2009fa} show a low value of $\lambda_{\max}\sim 0.365\mum$. It was suggested that the bulk of the polarization is produced by aligned grains within a single dense molecular cloud in the host galaxy ISM (\citealt{2008A&A...485L...9C}; \citealt{2013ApJ...779...38P}; \citealt{Patat:2015bb}). Therefore, for SNe 1986G and 2006X, in the following, we assume the entire extinction and polarization are produced by IS dust. {Although some small amount of CS dust is detected toward these SNe, its contribution is negligible in terms of the total reddening and polarization towards SNe.}

\subsection{SNe 2008fp and 2014J}
\subsubsection{Anomalous values of $R_{V}$ and $\lambda_{\max}$} 
SN 2008fp is located in the host galaxy ESO 428-G14, a peculiar galaxy with an active nucleus. Similar to SN 2006X, SN 2008fp exhibits an extreme value of $R_{V}\sim 1.2$ (\citealt{2013ApJ...779...38P}; \citealt{Cox:2014fq}). Its polarization data also demonstrate an anomalous value of $\lambda_{\max}\sim 0.15\mum$ (\citealt{Cox:2014fq}). It is believed that the extinction and polarization for SN 2008fp are mainly produced by IS dust due to the lack of variability in the absorption line profile and strong CN interstellar absorption features \citep{Cox:2014fq}.

SN 2014J is located in the host galaxy NGC 3034 (M82), which was discovered accidentally by a group of students \citep{2014CBET.3792....1F}. Photometric observations using Swift near-UV and optical-near IR data by \cite{2015ApJ...805...74B} show that the reddening of SN 2014J can be best fitted by the CCM law with $R_{V}=1.4$ (see also \citealt{2014ApJ...792..106W}; \citealt{2015MNRAS.453.3300A}). Polarimetric observations show an even more extreme value of $\lambda_{\max}=0.05\mum$ (\citealt{Patat:2015bb}; see also \citealt{Kawabata:2014gy}). 

{We note that the solid evidence for the peak wavelength is missing, and the values of $\lambda_{\max}$ reported in \cite{Patat:2015bb} are upper limits. Thus, the applicability of the empirical Serkowski law outside of the visible range is debatable. Nevertheless, with the goal of exploring dust properties and grain alignment for the wide range of grain size, especially for small grains that dominate UV extinction and polarization, the Serkowsli law will still be used for extrapolation due to the lack of real observations in the UV.}

\cite{Foley:2014br} {obtained similar fits for the two different models with and without CS dust. For the former model, they show} that the best-fit models for the extinction toward SN 2014J require equal reddening contributions from both CS and IS dust. Their best-fit parameters for the interstellar reddening is $R_{V}=2.59$ and $E_{B-V}=0.45\mag$ (i.e., $A_{V}=1.165\mag$), and CS dust contributes about the half extinction. On the other hand, appealing to the well-aligned polarization angles with the local spiral arms of the host galaxies and small variability before and after the maximum epoch, \cite{Patat:2015bb} argued that IS dust is a dominant contribution to the observed polarization, although adding the contribution of CS dust, they obtained more reasonable values of $R_{V}$ and $\lambda_{\max}$ for this case. {It is noted that the LOS to SN 2014J shows a multitude (>20) of absorption systems (revealed by the Ca, Na and K interstellar lines). As the host galaxy is seen almost edge-on, it is natural to expect that the LOS would intercept many intervening clouds. Unless we want a substantial fraction of those clouds (which have different velocities) to be very close to the SN, it is very hard to believe that up to 50$\%$ of the extinction is generated "locally" and the same applies to polarization.}

\subsubsection{Two-component extinction/polarization model}
\begin{figure*}\centering
\includegraphics[width=0.4\textwidth]{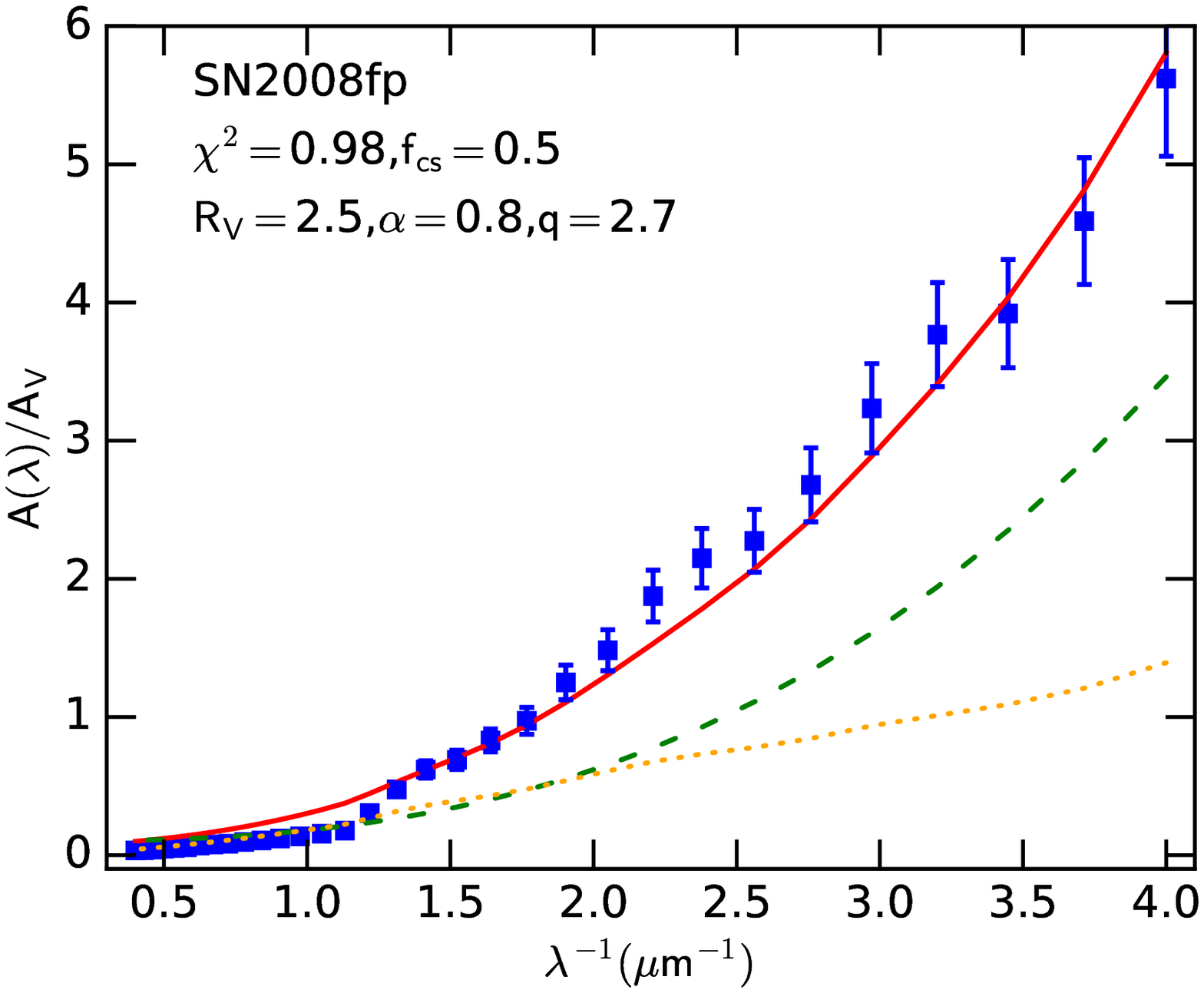}
\includegraphics[width=0.42\textwidth]{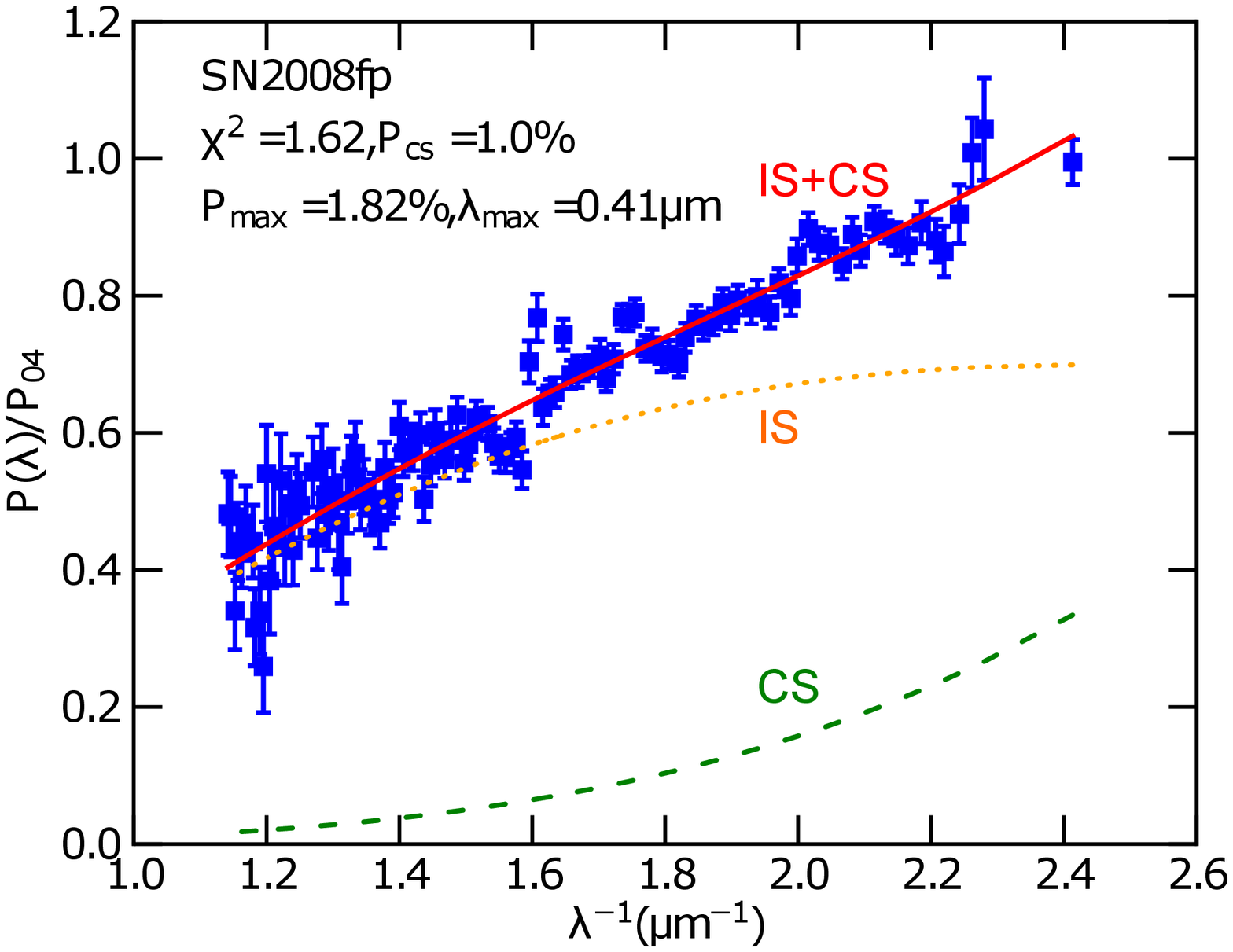}
\caption{Two-component extinction (left) and polarization (right) fits to the observed data for SN 2008fp. Data points are taken from \citep{Patat:2015bb}. Dotted and dashed lines represent the results for IS dust and CS dust, respectively, and solid lines show the total contribution from the two components.}
\label{fig:fit2}
\end{figure*}

{Due to uncertainty in the origin of anomalous $R_{V}$ and $\lambda_{\max}$ in SNe 2008fp and 2014J, we consider two models (model 1 and model 2). Model 1 assumes that the entire extinction and polarization are produced by IS dust (one dust component), whereas model 2 considers two dust components (IS and CS dust) contributing to the observed data. 

The observed extinction from both IS and CS dust is given by
\bea
A_{\obs}(\lambda) = A_{\rm is}(\lambda) + A_{\rm cs}(\lambda) ,\label{eq:Aobs}
\ena
where $A_{\rm is}(\lambda)$ is given by the CCM law, and the latter is a power law \citep{2008ApJ...686L.103G}:
\bea
\left(A(\lambda)/A_{V}\right)_{\rm cs} = 1 -\alpha + \alpha\left(\frac{0.55\mum}{\lambda} \right)^{q}, \label{eq:ACSM}
\ena
where $\alpha$ and $q$ are free parameters.

Let $f_{\rm cs}$ is the fraction of the total observed visual extinction $A_{V}$ contributed by CS dust. Then, we can obtain the followings
\bea
A_{\rm cs}(\lambda) = f_{\rm cs}A_{V}\times \left(A(\lambda)/A_{V}\right)_{\rm cs},\\
A_{\rm is}(\lambda) =  (1-f_{\rm cs})A_{V}\times \left(A(\lambda)/A_{V}\right)_{\rm is}.
\ena

Similarly, including the polarization due to Rayleigh scattering by CS dust (see \citealt{2013ApJ...775...84A} for details), the observed polarization is then equal to
\bea
P_{\rm obs}(\lambda) = P_{\rm is}(\lambda) + P_{\rm cs}\left(\frac{\lambda}{0.4\mum}\right)^{-4},\label{eq:pISM_pR}
\ena
where $P_{\rm is}(\lambda)$ is the polarization from IS dust (Eq. \ref{eq:serkowski}), and $P_{\rm cs}$ denotes the polarization due Rayleigh scattering. {Above, we have adopted the $-4$ slope for the polarization by Rayleigh scattering, presumably due to Mie scattering by small grains and molecules. Yet, the precise slope of Rayleigh scattering is uncertain, depending on the source geometry and observation wavelength. Moreover, we have assumed that the interstellar polarization and scattering polarization is additive, which is only valid when the polarization directions from the two processes are aligned. There might be some special geometry of CS dust cloud that can satisfy these condition, which is suggested in Section \ref{sec:model}) to explain the constant polarization angles toward SN 2014J.}

To infer the best-fit model parameters $\alpha, q, f_{\rm cs}$, and $R_{V}$, we use the publicly available package lmfit-py\footnote{http://cars9.uchicago.edu/software/python/lmfit/index.html} to fit $A_{\obs}(\lambda)$ to the observed extinction. Similarly, we fit $P_{\obs}(\lambda)$ to the observed polarization to find $P_{\rm cs}, \lambda_{\max}, P_{\max}$, and $K$. The error for the extinction are assigned to be 10 percent, and the polarization fit is interpolated from the real observational data.

Our fits using the two-component model return reasonable parameters with $R_{V}\approx 2.2$ for IS dust contributing $A_{V}=0.29$ to the total observed extinction, $\lambda_{\max}=0.41\mum$, and $K=1.03$ (see details in Figure \ref{fig:fit2} for SN 2008fp). We also rerun for SN 2014J and obtain a good fit with the comparable parameters as in previous works (\citealt{Foley:2014br};\citealt{Patat:2015bb}).}

\section{Dust Model and Observational Constrains}\label{sec:theory}
\subsection{Dust Model}
We adopt a mixed-dust model consisting of astronomical silicate grains and graphite grains (see \citealt{2001ApJ...548..296W} ; \citealt{2007ApJ...657..810D}). Because graphite grains are unlikely aligned with the magnetic field (\citealt{2006ApJ...651..268C}; see LAH15 for a review), we assume that only silicate grains are aligned while carbonaceous grains are randomly oriented. Oblate spheroidal grains with axial ratio $r=2$ are considered.

The extinction of starlight induced by randomly oriented grains in units of magnitude is given by
\bea
A({\lambda})&= \sum_{j=\sil,\carb} 1.086\int\int C_{\ext}^{j}(a)\left(\frac{dn^{j}}{da}\right)dadz,\label{eq:Aext}
\ena
where $a$ is the effective grain size defined as the radius of an equivalent sphere of the same grain volume, $dn^{j}/da$ is the grain size distribution of dust component $j$, $C_{\ext}$ is the extinction cross-section, and the integration is taken along the entire line of sight $z$.

The degree of polarization (in unit of $\%$) of starlight due to differential extinction by aligned grains along the line of sight is computed by
\bea
P({\lambda})= \sum_{j=\sil} 100\int \int \frac{1}{2}C_{\pol}^{j}(a)f^{j}(a)
\frac{dn^{j}}{da}dadz,\label{eq:Plam}
\ena
where $C_{\pol}$ is the polarization cross-section, and $f^{j}(a)$ is the effective degree of grain alignment for dust component $j$ with size $a$ (hereafter alignment function). Here we take $C_{\ext}$ and $C_{\pol}$ computed for different grain sizes and wavelengths from \cite{2013ApJ...779..152H}.

\subsection{Observational Constraints}
\subsubsection{Gas-to-dust ratio}
The important constraint of the dust model is the gas-to-dust mass ratio, which ensures that the total gas mass relative to the total dust mass must be constrained by observations. The gas-to-dust mass ratio is usually represented through $N_{\H}/A_{\V}$ where $N_{\H}$ is the gas column density. For the Galaxy, measurements give $ N_{\H}/E_{B-V}=5.8\times 10^{21}\cm^{-2}\mag^{-1}$, which corresponds to $N_{\H}/A_{V}=5.8\times 10^{21} \cm^{-2}\mag^{-1}/R_{V}$.

The gas-to-dust mass ratio for the SN 1986G hosted galaxy is measured by Herschel (\citealt{2012MNRAS.422.2291P}) and SINGS (\citealt{2007ApJ...663..866D}). The former provides an average gas-to-dust mass ratio to be $103\pm 8$, which is similar to our Galaxy. Therefore, we can take the same ratio of the Galaxy, $N_{\H}/E_{B-V}=5.8\times 10^{21}\cm^{-2}\mag^{-1}$, for the LOS to SN 1986G with $E_{B-V}=0.79\mag$. It is worth mentioning that \cite{1989A&A...215...21D} estimated $N_{\H}\sim 5\times 10^{21}\cm^{-2}$ for $E_{B-V}=0.9\pm 0.1 \mag$.

The reddening to SN 2006X is estimated to be $E_{B-V}=1.42\pm 0.4 \mag$, assuming the Galactic foreground equal to $0.026$ mag. Observations for NGC 4321 in \cite{2007ApJ...663..866D} show that the dust-to-gas ratio is about two times higher than that of our Galaxy with metallicity comparable to the Solar abundance. Therefore, for SN 2006X, we assume $N_{\H}/A_{V}={5.8\times 10^{21} \cm^{-2}\mag^{-1}}/2{R_{V}}$, or $N_{\H}/E_{B-V}=2.9\times 10^{21}\cm^{-2}\mag^{-1}$ for $R_{V}=1.31$. 

For SN 2008fp, \citealt{Cox:2014fq} estimated $N(\H_{I})=6.2_{-2.1}^{+3.2}\times 10^{20}\cm^{-2}$, and $N(\H_{2})=7.2_{-2.7}^{+4.8}\times 10^{20}\cm^{-2}$, which gives $N_{\H}=N(\H_{I})+2N(\H_{2})=2.1\times 10^{21}\cm^{-2}$. For model 1, we get $N_{\H}/A_{V}=2.95\times 10^{21}\cm^{-2}\mag^{-1}$ for $A_{V}=0.71\mag$. For $R_{V}=1.2$, it yields $N_{\H}/E_{B-V}=3.52\times 10^{21}\cm^{-2}\mag^{-1}$, indicating that the dust-to-gas ratio in this galaxy is higher than that in the Galaxy. {For model 2 in which the scattering by IS dust is included, the ratio becomes $N_{\H}/E_{B-V}=1.6\times 10^{22}\cm^{-2}\mag^{-1}$ for $E_{B-V}=0.13\mag$.}

For SN 2014J, the gas column density is estimated to be ${\rm log} N_{\H}=21.53\pm 0.13$ (see Table 3 in \citealt{Ritchey:2014uq}). Thus, $N_{\H}/E_{B-V}=2.47\times 10^{21}\cm^{-2}\mag^{-1}$ for model 1 with $E_{B-V}=1.37\mag$. For model 2 in which IS dust only contributes $E_{B-V}=0.45$ \citep{Foley:2014br}, we obtain $N_{\H}/E_{B-V}=7.52\times 10^{21}\cm^{-2}\mag^{-1}$.

In Table \ref{tab:SN} we summarize the parameters of four SNe Ia used for inversion problem (see also \citealt{Patat:2015bb}). {Again, we note that the adopted values of $\lambda_{\max}, K$ taken from \cite{Patat:2015bb}, which were derived from the fits to observations using the Serkowski law, may not reflect underlying physics adequately. However, we conservatively assume that the ISM of the hosted galaxies can be characterized by these peculiar values of $\lambda_{\max}$ and $R_{V}$ and explore the properties and grain alignment of interstellar dust that can successfully reproduce such peculiar values.}  

\begin{table*}
\centering
\caption{Physical parameters for SNe Ia }\label{tab:SN}
\begin{tabular}{l l l l l l l l l l l }\hline\hline\\
SNe & Host & Host type & $\lambda_{\max}(\mum)$ & $p_{04}(\%)$ & $K$ & $R_{V}$ & $A_{V}(\mag)$ & $E_{B-V}(\mag)$   & $N_{\H}/E_{B-V}$\cr
& & & & & & & & & $(\cm^{-2}\mag^{-1})$\cr\hline
1986G & NGC~5128 & starburst &  0.43 & 5.1  & 1.15  & $2.57$  & $2.03$  & 0.79  & 5.8E+21\cr

2006X & NGC~4321 &  normal & 0.36  & 7.8  & 1.47  & $1.31$  & 1.88  & 1.44  & 2.9E+21\cr

2008fp (mod 1) & ESO~428-G14 &  Seyfert & 0.15   & 2.6  & 0.40  & $1.20$  & $0.71$  & 0.59  & 3.5E+21\cr
2008fp (mod 2) & -- &  -- & 0.41  & 1.8  & 1.03  & $2.20$  & $0.29$  & 0.13  & 1.6E+22\cr

2014J (mod 1) & NGC~3034 &  starburst  &  0.05   & 6.6  & 0.40    & 1.40   & 1.85   & 1.37  & 2.5E+21\cr
2014J (mod 2) & -- & --  &  0.35  & 3.7  & 1.15   & 2.59   & 1.17   & 0.45  & 7.5E+21\cr

\cr
\hline
\cr
\end{tabular}
\end{table*}

\subsubsection{Metal abundances in galaxies}

One important constraint is the fraction of metal elements (e.g., Si, Fe, Mg, and C) incorporated into dust. Let us now evaluate the silicate volume per H atom. Assuming that all Mg, Si and Fe budget (\citealt{1989GeCoA..53..197A}) are incorporated in the solid silicate material, then, for each Si atom, we would form a structure Mg$_{1.1}$Fe$_{0.9}$SiO$_{4}$. The volume of silicate material per H atom is $V_{\sil,0}/\H = ({\rm Si}/\H)m_{\sil}/\rho_{\sil}=2.57\times 10^{-27}\cm^{3}/\H$ for $\rho_{\sil}=3.5\g\cm^{-3}$. Similarly, for graphite with $\rho_{\gra}=2.2\g\cm^{-3}$, we get $V_{\gra,0}=2.23\times 10^{-27}\cm^{3}/\H$ with ${\rm C}/\H=2.46\times 10^{-4}$. The volume of $j$ dust component of the model is evaluated by $
V_{j}=\int \left(4\pi a^{3}/{3}\right) {dn_{j}}/{da} $. 

\section{Monte Carlo Inversion Technique and Results}\label{sec:method}

Inversion technique has frequently been used to infer the grain size distribution of dust grains in the ISM of the Milky Way (\citealt{1994ApJ...422..164K};  \citealt{2000ApJ...532.1021L}; \citealt{2004ApJS..152..211Z}), and in nearby galaxies (e.g, small Magellanic cloud \citep{2003ApJ...588..871C}. \cite{Draine:2009p3780} used Levenberg-Marquart method to infer both the grain size distribution and alignment function of interstellar grains in the Galaxy characterized by the typical values of $R_{V}=3.1$ and $\lambda_{\max}=0.55\mum$. Here we follow \cite{{2013ApJ...779..152H},{2014ApJ...790....6H}} (hereafter HLM13, HLM14) using the Monte Carlo method to find best-fit grain size distribution and alignment function for interstellar grains in the SNe Ia hosted galaxies with anomalous extinction and polarization data. 

\subsection{Simulation-based Inversion Technique}

To study dust properties and grain alignment toward SNe Ia, we will find the best-fit grain size distribution and alignment function by fitting the observed extinction and polarization curves with the theoretical models $A_{\rm mod}$ and $p_{\rm mod}$. The observed data for the SNe Ia are calculated using Equation (\ref{eq:serkowski}) and the CCM law with the relevant parameters shown in Table \ref{tab:SN}. Our nonlinear least square fitting code uses Monte Carlo simulation method to minimize an objective function $\chi^{2}=\chi^{2}_{\rm mod}+\chi^{2}_{\rm con}$ where $\chi^{2}_{\rm mod}=\chi_{\ext}^{2}+\chi^{2}_{\pol}$ describes the misfit between the model extinction and polarization and the observed data, and $\chi^{2}_{\rm con}$ describes the model constraints, including the smoothness of $dn/da$ and the monotonic variation of $f_{a}$ (see HLM13, HLM14 for details).

The important constraint for the polarization model and the alignment function $f(a)$ is that, for the maximum polarization efficiency $p_{\max}/A(\lambda_{\max})=3\% \mag^{-1}$ (see \citealt{2003ARA&A..41..241D} for a review), we expect that the conditions for grain alignment are optimal, which corresponds to the case in which the alignment of big grains can be perfect, and the magnetic field is regular and perpendicular to the line of sight. Thus, we set $f(a=a_{\max})=1$. For a given line of sight with lower $p_{\max}/A(\lambda_{\max})$, the constraint $f(a=a_{\max})$ should be adjusted such that $f(a=a_{\max})=(1/3)p_{\max}/A(\lambda_{\max})$. For big grains, we expect the monotonic increase of $f(a)$ versus $a$, thus a monotonic constraint is introduced for $a>50$\AA. For $a<50$\AA, expecting a different alignment mechanism based on resonance paramagnetic relaxation that results in a peaky alignment function (HLM13), we don't constrain the monotonic decrease for this very small population. Other constraints include the non-smoothness of $dn/da$ and $f(a)$ (see \citealt{2006ApJ...652.1318D}). 

Because the extinction and polarization data in far-UV wavelength ($\lambda<0.25\mum$) toward the considered SNe Ia are unavailable, we will not attempt to invert the data for the far-UV wavelengths, which is mainly contributed by ultrasmall grains (including polycyclic aromatic hydrocarbons). Thus, we consider $\lambda=0.25-2.5\mum$ and compute the extinction and polarization model given by Equations (\ref{eq:Aext}) and (\ref{eq:Plam}), respectively).  We use 64 bins of grain size in the range $a=10$\AA~to $1\mum$ and 64 bins of the wavelength. 

The fitting procedure is started with an initial size distribution $n(a)$ that best reproduces the observational data for the typical ISM (model 3 in \citealt{Draine:2009p3780}) and is iterated until the convergence criterion is satisfied. At each iteration step, for each size bin, we generate $32$ independent sample models by perturbing all model parameters $n_{a}, f_{a}$ using the Gibbs sampling algorithm, and retain the best model determined by the current minimum $\chi^{2}$. The process is looped over all size bins and the final minimum $\chi^{2}$ is obtained for which a unique best-fit model is retained. An iteration step is accomplished. The convergence criterion is based on the decrease of $\chi^{2}$ after one step: $\epsilon=(\chi^{2}(n,f)-\chi^{2}(\tilde{n},\tilde{f}))/\chi^{2}(n,f)$. If $\epsilon\le \epsilon_{0}$ with $\epsilon_{0}$ sufficiently small, then the convergence is said to be achieved (see HLM13, HLM14). With the value $\epsilon_{0}=10^{-3}$ adopted, the convergence is slow for some lines of sight, we stop the iteration process when the variation of $\chi^{2}$ is relatively stable.

\subsection{Inversion Results}\label{sec:result}
\subsubsection{Convergence and Best-fit models}

\begin{figure*}
\centering
\includegraphics[width=0.42\textwidth]{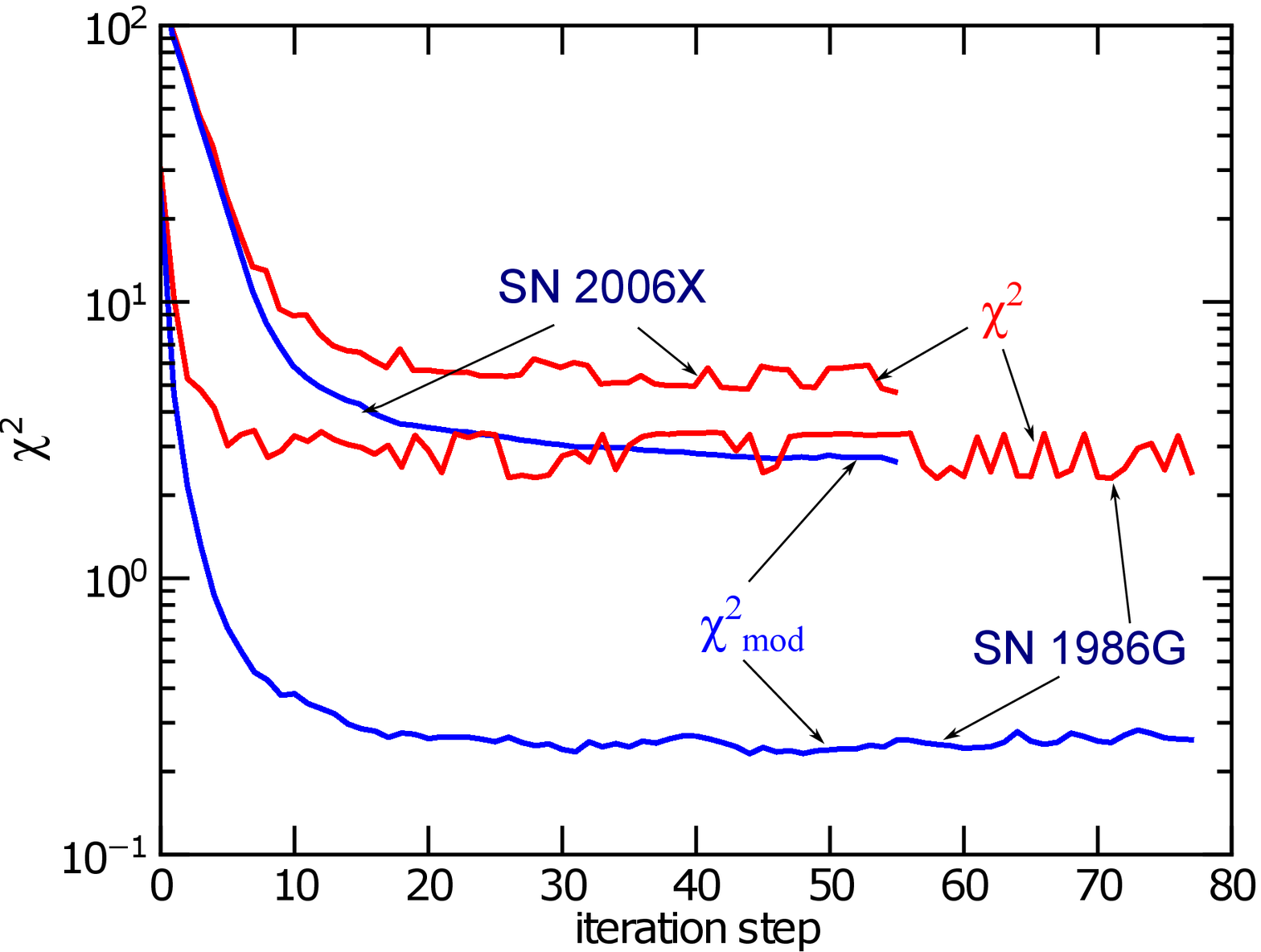}
\includegraphics[width=0.42\textwidth]{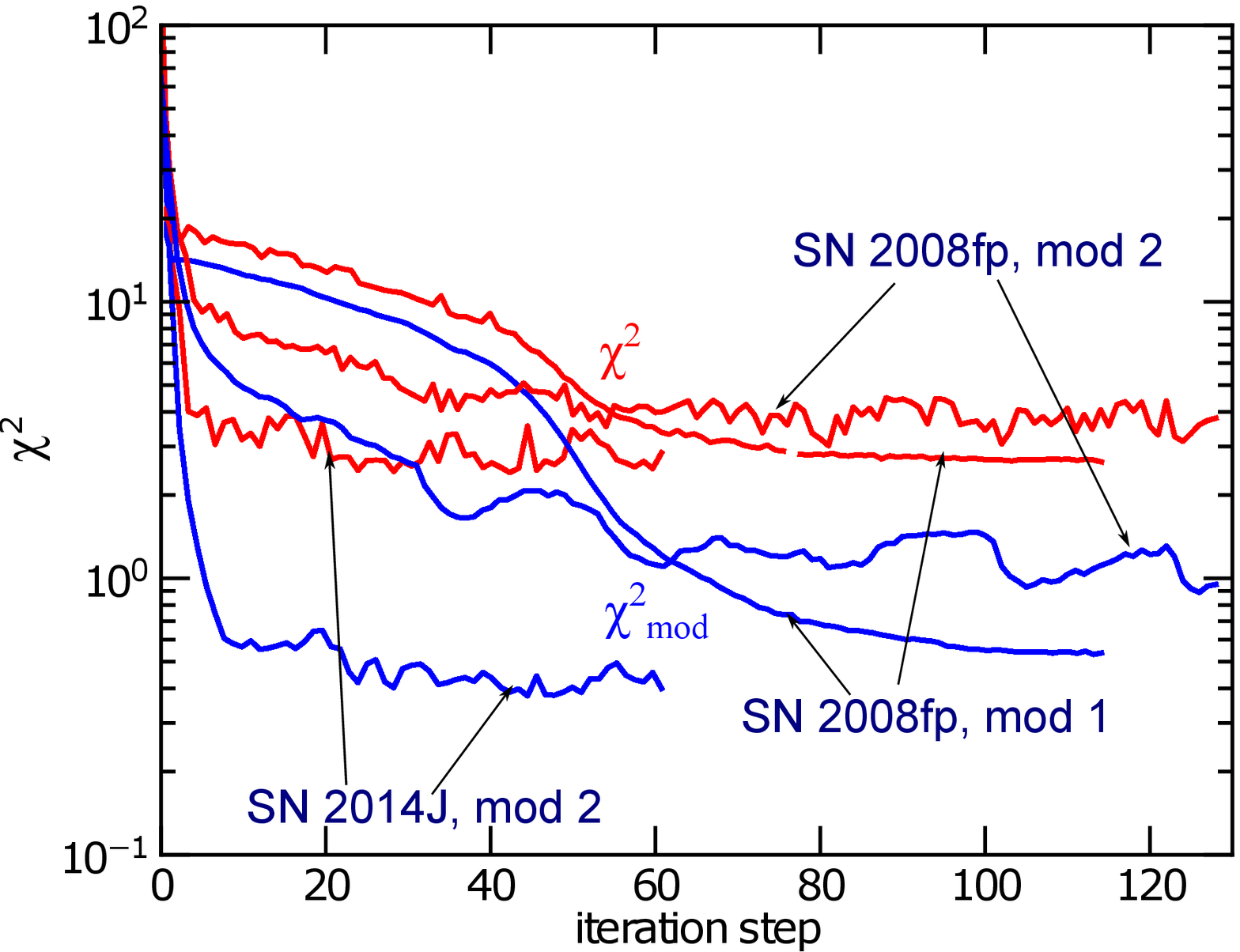}
\caption{Reduction of total $\chi^{2}$ and $\chi^{2}_{\rm mod}$ (i.e., mistfit between the model and data only) vs. the iteration step for the SNe Ia considered. A fast steady reduction in $\chi^{2}$ is achieved after about 30 steps for all cases except SN 2008fp.}
\label{fig:chisq}
\end{figure*}

For SNe 1986G, 2006X, we run the inversion code assuming the entire extinction and polarization are produced by the IS dust, and obtain a gradual reduction of $\chi^{2}$. For SNe 2008fp and 2014J, we run the inversion code for both model 1 and model 2 (see Table \ref{tab:SN}). For SN 2014J (model 1), our simulations do not show the reduction of $\chi^{2}$ and the minimum $\chi^{2}$ obtained is too large (i.e., $\chi^{2}\sim 10^{2}$), which indicates that the inversion fails to converge. For model 2, our simulations exhibit a good reduction of $\chi^{2}$. Detailed evolution of $\chi^{2}$ and $\chi_{\rm mod}^{2}$ vs. the iteration step for four SNe Ia are shown in Figure \ref{fig:chisq}.

From the figure, it can be seen that the achieved goodness of fits is excellent for SN 1986G and SN 2014J (model 2), which have terminal $\chi_{\rm mod}^{2} \sim 0.3-0.4$. For the case with extreme $R_{V}$, SN 2006X, the goodness of fit is poorer with $\chi_{\rm mod}^{2}\sim 3$. SN 2008fp (model 1) with both extreme $R_{V}$ and $\lambda_{\max}$ has rather slow reduction in $\chi^{2}$, which requires $\sim 100$ iteration steps to reach the stable value of $\chi^{2}$. Interestingly, SN 2008fp (model 2) shows similar trend as model 1, but reaches higher $\chi^{2}$ after $\sim $100 steps.

\begin{figure*}
\centering
\includegraphics[width=0.4\textwidth]{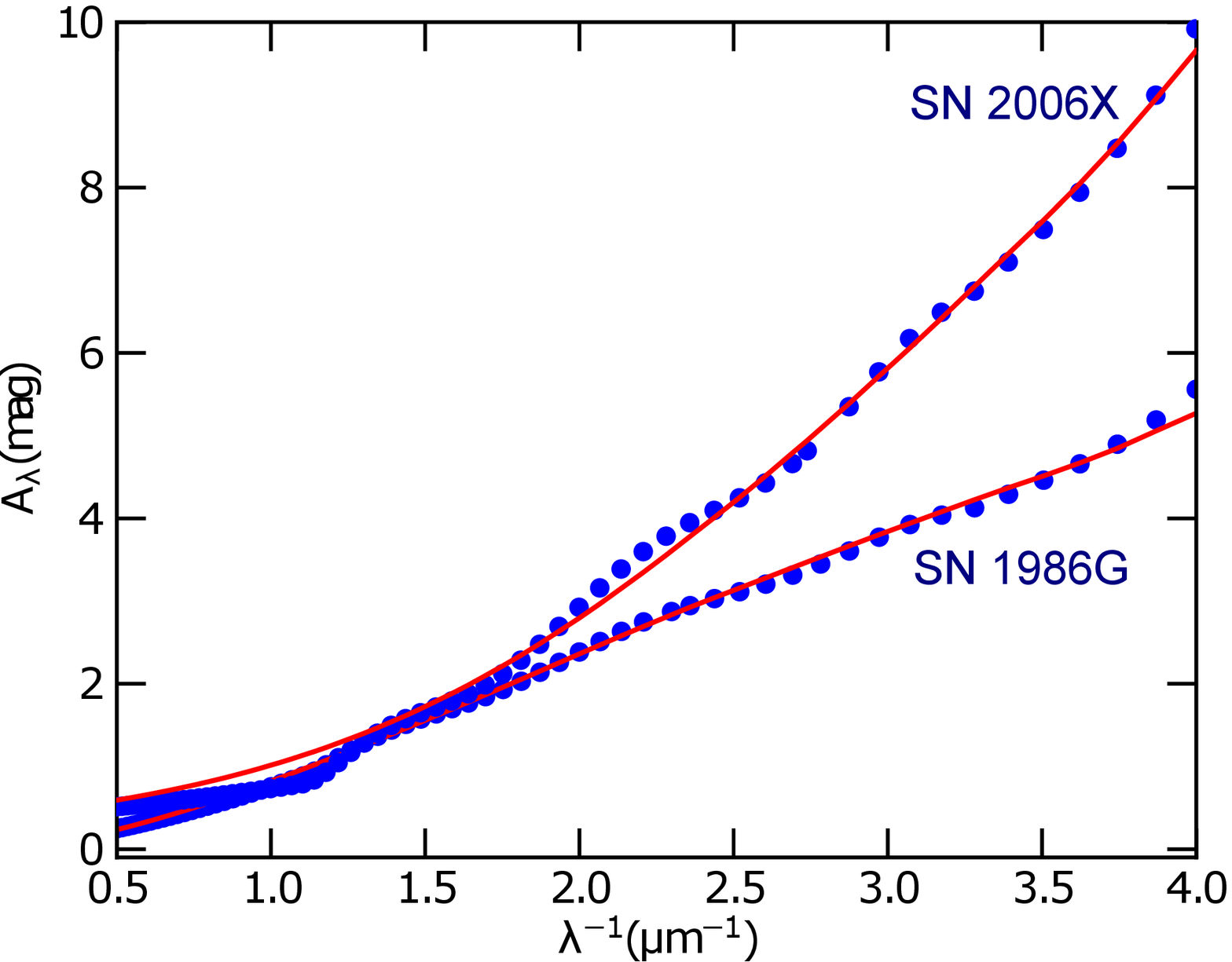}
\includegraphics[width=0.4\textwidth]{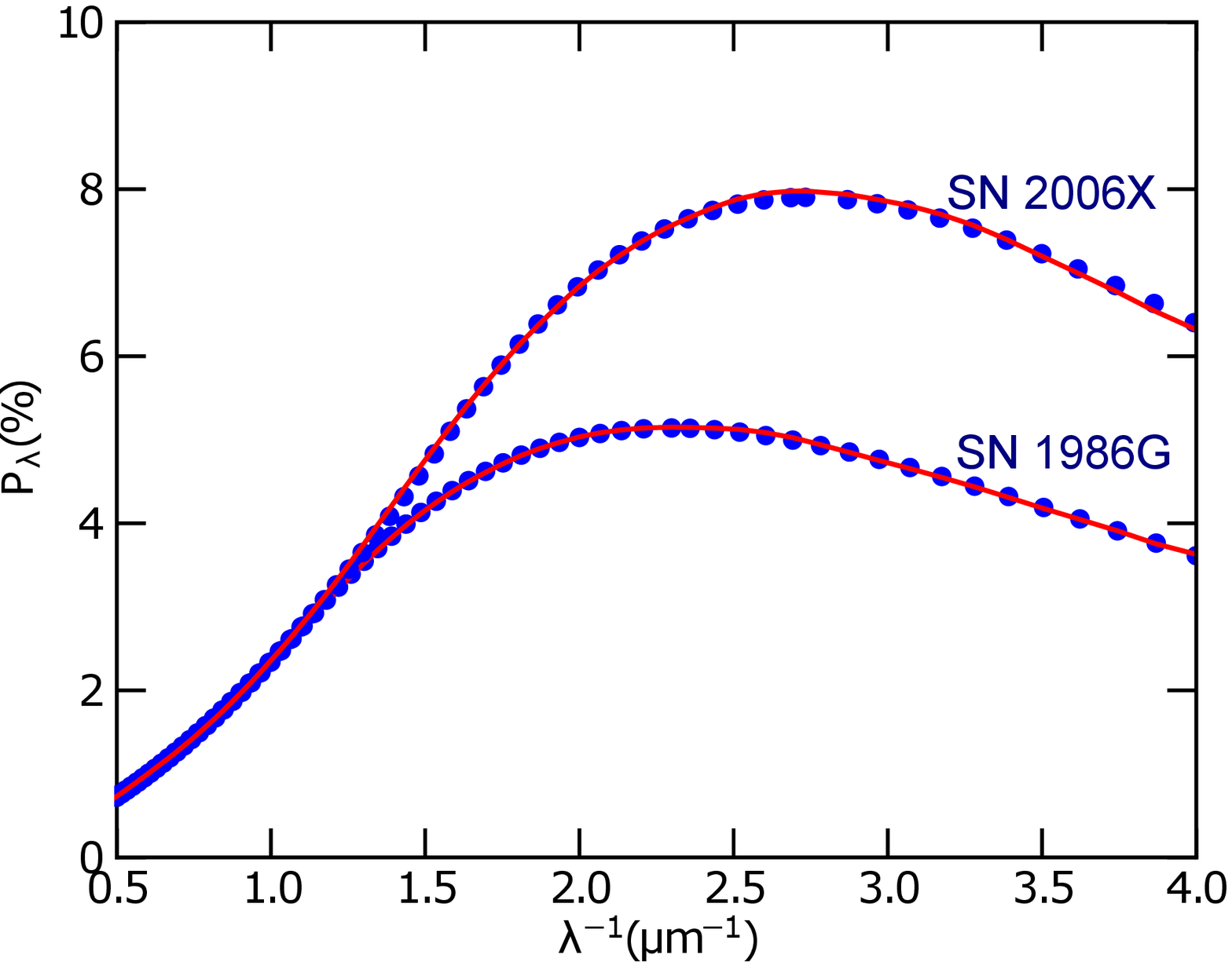}
\includegraphics[width=0.4\textwidth]{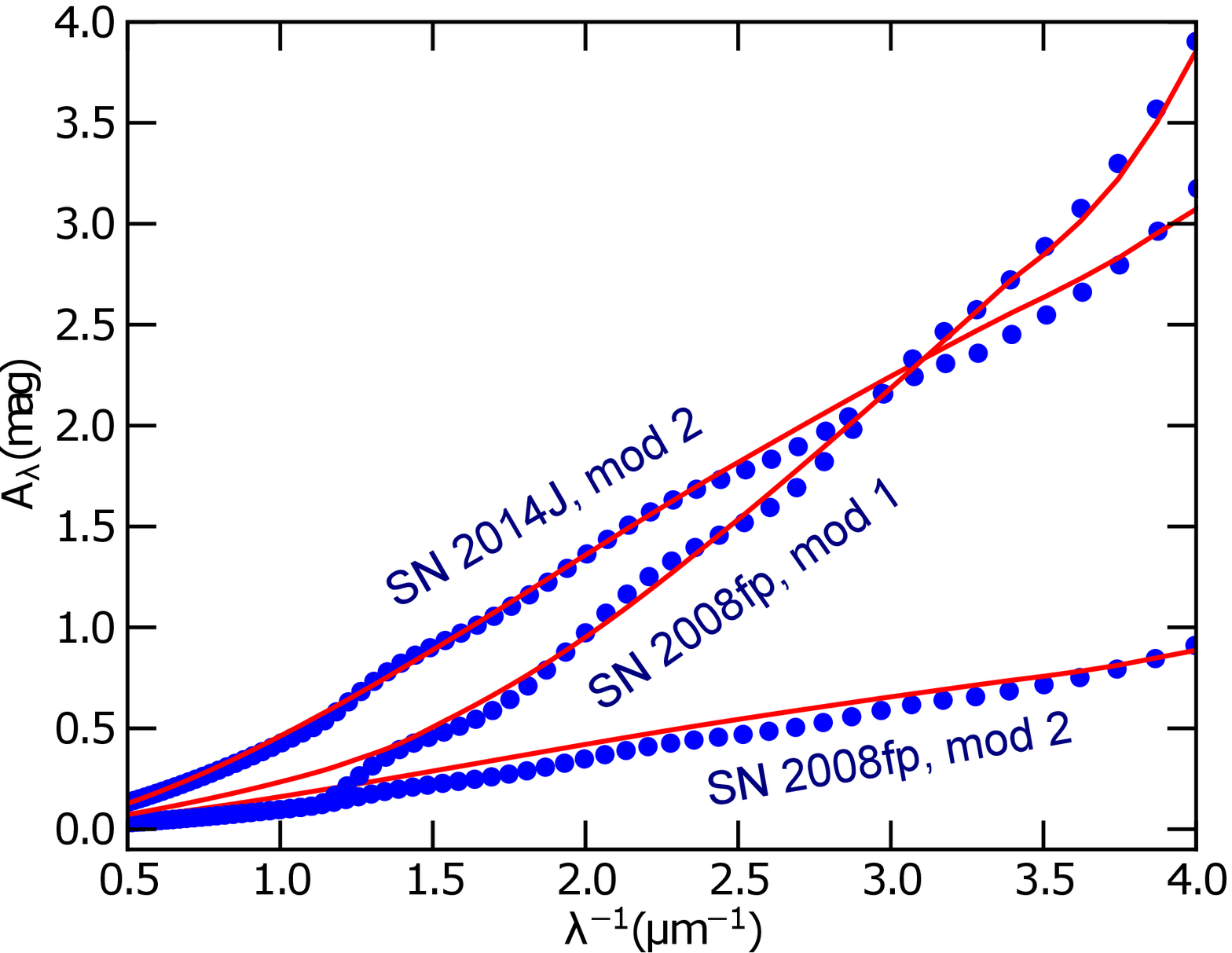}
\includegraphics[width=0.4\textwidth]{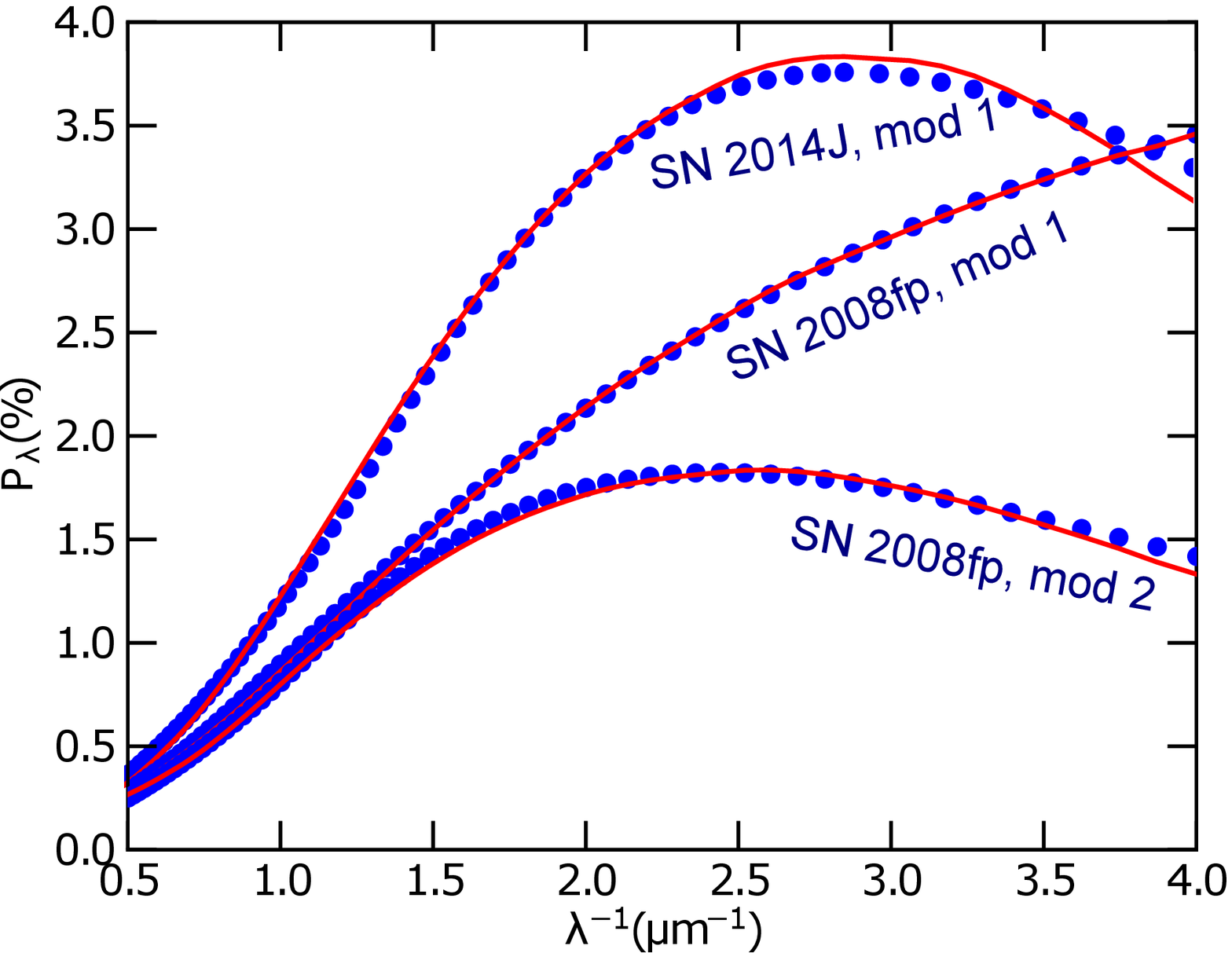}
\caption{Best-fit models (solid lines) vs. observed data (circles) for extinction (left) and polarization (right) of SNe 1986G and 2006X. Excellent fit is obtained both for the polarization and extinction curves.}
\label{fig:polmod2}
\end{figure*}

Figure \ref{fig:polmod2} shows best-fit models to the observed extinction (left panels) and polarization curves (right panels) for SNe 1986G and 2006X, SNe 2008fp, and 2014J. 

\subsubsection{Best-fit grain size distribution and alignment function}

\begin{figure*}
\centering
\includegraphics[width=0.4\textwidth]{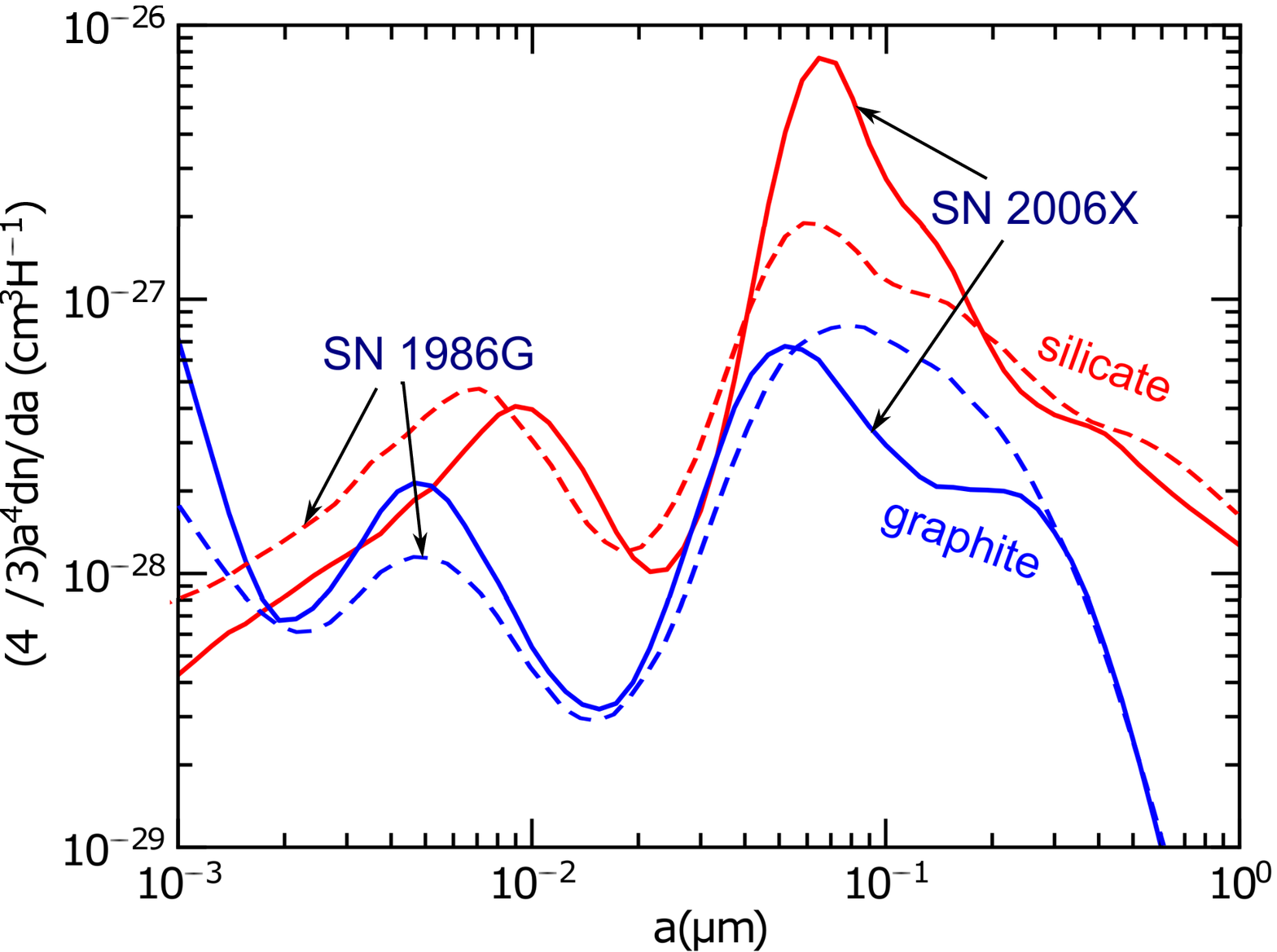}
\includegraphics[width=0.4\textwidth]{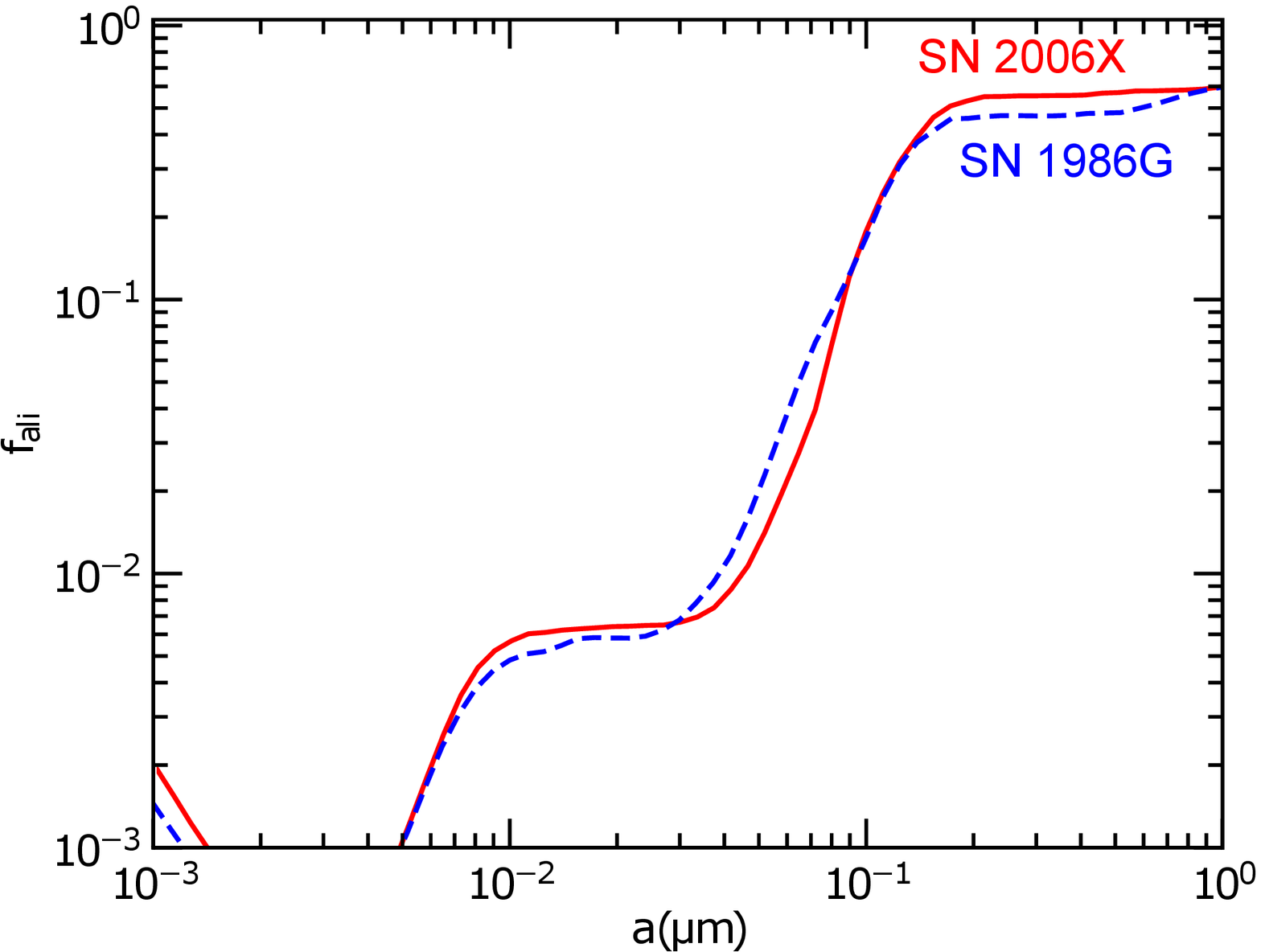}
\includegraphics[width=0.41\textwidth]{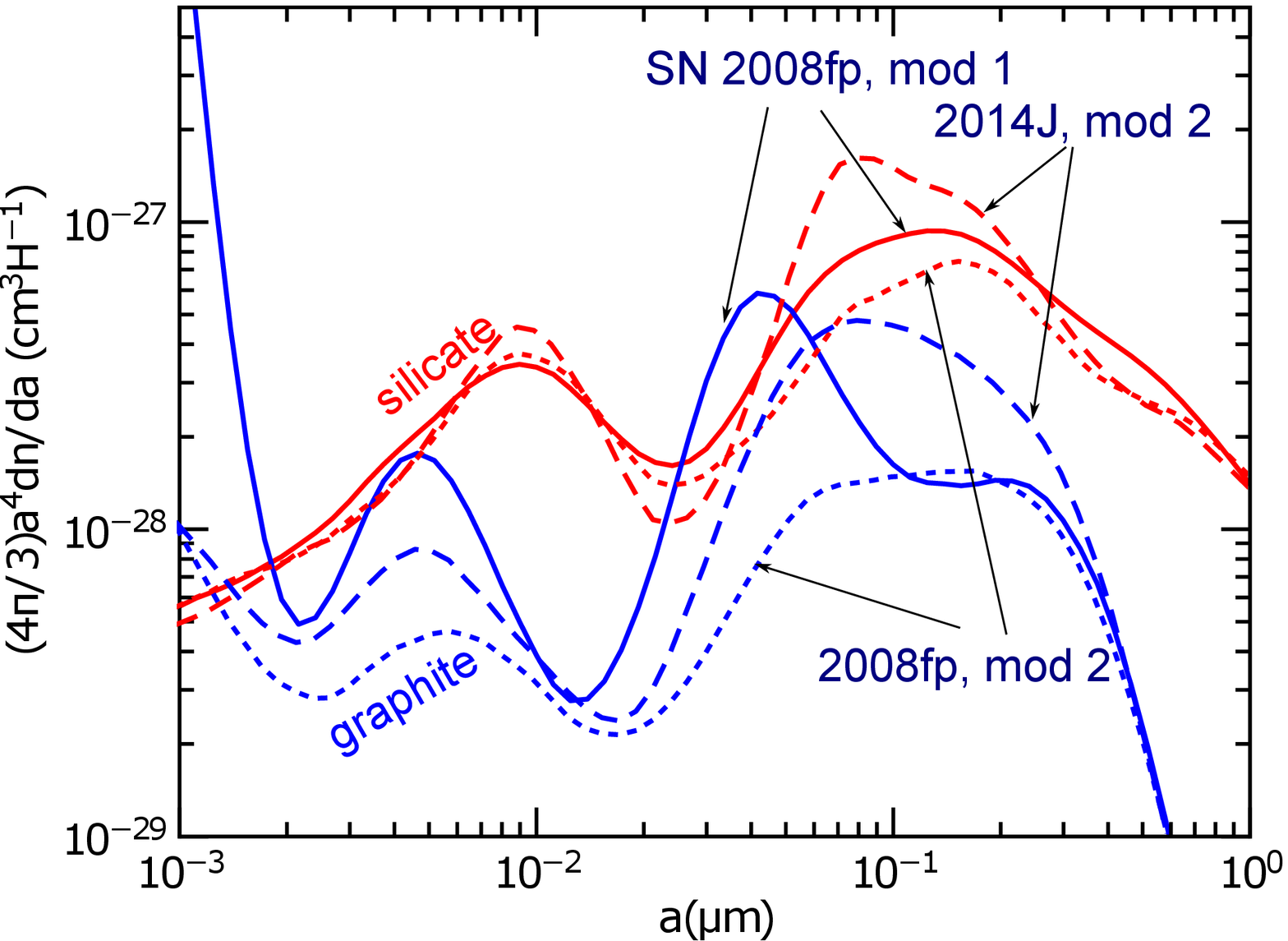}
\includegraphics[width=0.4\textwidth]{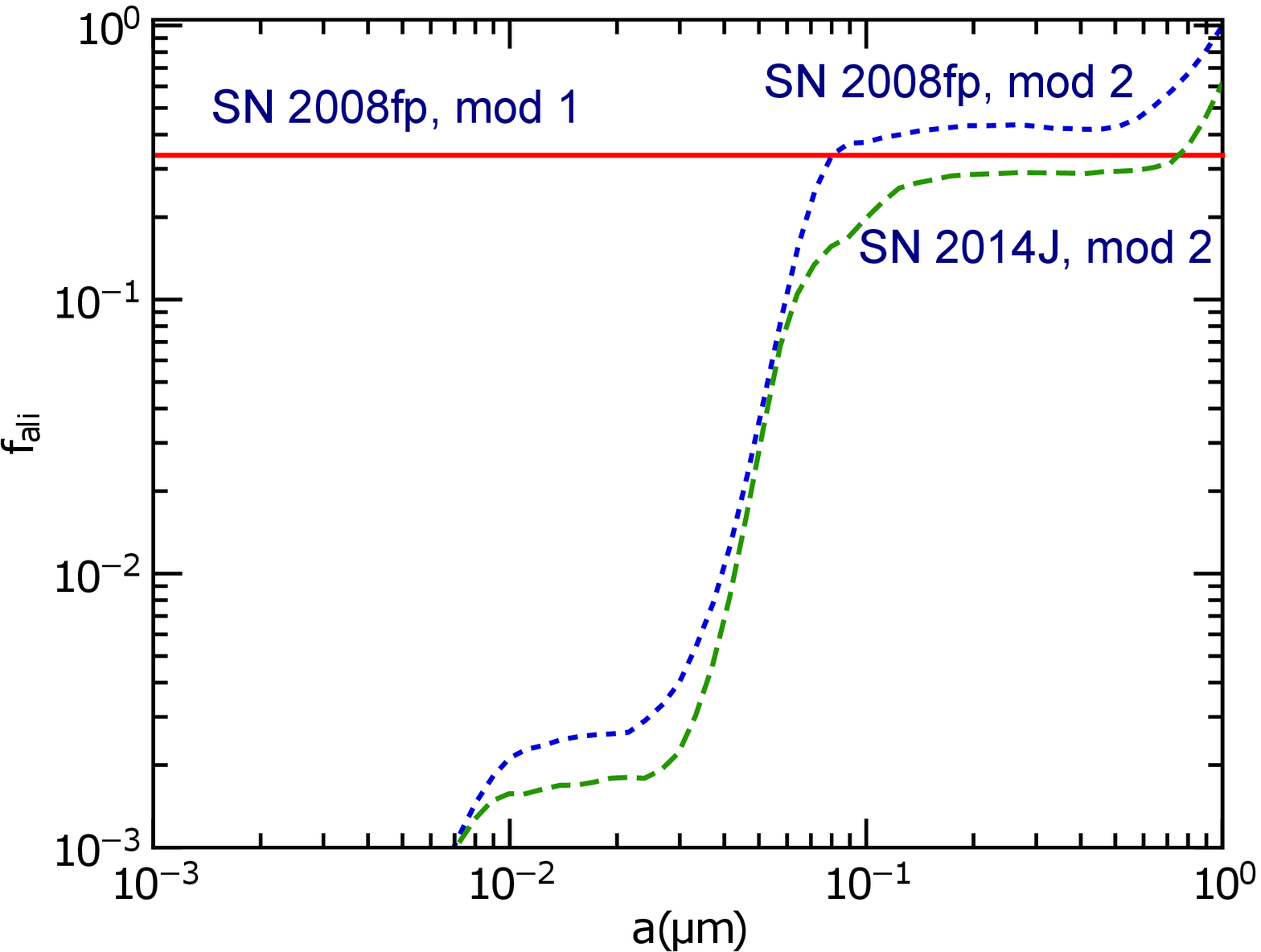}
\caption{Grain mass distribution (red for silicate and blue for graphite) and alignment function for the best-fit models for the considered SNe Ia. SN 2006X exhibits a substantial enhancement in the mass of small grains $a\sim 0.06\mum$ compared to SN 1986G. Three SNe 1986G, 2006X, 2008fp (model 2) and 2014J (model 2) have similar alignment functions with low alignment degree of small grains of $a<0.04\mum$. The alignment function of SN 2008fp (model 1) appears to be peculiar.}
\label{fig:nafali1}
\end{figure*}

Figure \ref{fig:nafali1} shows the best-fit model parameters for our considered SNe Ia, including the grain mass distribution and alignment function. The best-fit mass distributions show the lack of large grains of $a>0.1\mum$. For instance, compared to SN 1986G, SN 2006X with a much lower value of $R_{V}$ requires a significant increase in the mass of small silicate grains at $a\sim 0.06\mum$. In particular, both SNe 1986G and 2006X have rather similar best-fit alignment functions due to their similar $\lambda_{\max}$ {inferred from fitting the Serkowski law to the observation data}. In addition, the alignment of small grains $a\sim 0.05-0.1\mum$ is significantly increased compared to the typical alignment function in the Galaxy with $\lambda_{\max}\sim 0.55\mum$ (e.g., see HLM14), which is required to reproduce the low values of  $\lambda_{\max}\sim 0.35-0.45\mum$.

Figure \ref{fig:nafali1} (lower panels) shows the results for SNe 2008fp (models 1 and 2) and 2014J (model 2). Compared to SN 2006X with similar, low value of $R_{V}$, SN 2008fp (model 1) requires a higher mass of very small silicate grains at $a<0.010\mum$ but lower mass of small silicate grains $a\sim 0.06\mum$. Similarly, the mass of small graphite grains must be enhanced to compensate for the reduction of small silicate in order to reproduce the same $R_{V}$ as SN 2006X. Interestingly, for SN 2008fp (model 1), the alignment degree is independent on the grain size, in which small grains must be aligned as efficient as large grains. The best-fit size distribution and alignment function obtained for SN 2014J (model 2) are reminiscent of SN 1986G due to their similar low values of $R_{V}$ and {inferred values of} $\lambda_{\max}$.

\section{Discussion}\label{sec:dis}
\subsection{Dust properties toward SNe Ia: enhancement in the mass of small grains}
In this paper, we have applied our inversion technique to infer a specific model of interstellar grains in the distant galaxies probed by SNe Ia that posses anomalous observational data. To reproduce the anomalous extinction and polarization for four SNe Ia (1986G, 2006X, 2008fp, and 2014J), our best-fit models indicate that dust grains essentially have enhanced mass at small sizes ($a< 0.1\mum$), whereas large grains ($a>0.1\mum$) are subdominant. This is essentially in agreement with previous works (\citealt{2008ApJ...677.1060W}; \citealt{2013ApJ...779...38P}; \citealt{Patat:2015bb}). 

In particular, SN 2006X with an extreme value of $R_{V}=1.31$ but a low value of $\lambda_{\max}=0.365\mum$ requires a largest enhancement in the mass of small grains of size $a\sim \lambda_{\max}/2\pi \sim 0.06\mum$ among the considered SNe Ia. This is easily understood because to reproduce the extreme value $R_{V}$, it requires an increase in mass of small grains, but such additional mass must be well concentrated at not very small size to reproduce a not very low value of $\lambda_{\max}$. The mass fraction of graphite in this model is estimated to be $q_{\gra}\sim 8\%$, about 3 times lower than that in the Galaxy ($q_{\gra}\sim 20\%$), which indicates the {decrease of the} depletion of C into the grain. This result appears to be in agreement with the high abundance of gas phase C (about 2 times higher than for the Galaxy) reported by \cite{Cox:2014fq} based on the analysis of CN absorption lines from the dense cloud.

For SN 2008fp that exhibits unusually low values of both $R_{V}$ and $\lambda_{\max}=0.148\mum$ (model 1 without CS dust), our inversion results reveal a high mass fraction of carbonaceous grains (e.g, at $a\sim 0.04\mum$) and an efficient alignment of grains of all sizes. When including Rayleigh scattering by CS dust (model 2), we derived more reasonable parameters for the extinction and polarization curves (i.e., $R_{V}=2.2$ and $\lambda_{\max}=0.42\mum$). As a result, our best-fit results do not reveal special grain alignment function as in model 1. This model also requires a lower mass fraction of graphite, $q_{\gra}\sim 11\%$, which seems to be consistent with estimates in \cite{Cox:2014fq}.
 
For SN 2014J, our inversion for model 1 is unsuccessful, i.e., we cannot reach a target $\chi^{2}$ (e.g., $\chi^{2}<10$). The inversion for model 2 is successful and provides a good fit to the observational data. The best-fit grain size distribution is similar to that for SN 1986G, which is expected due to their similar values of $R_{V}$ and $\lambda_{\max}$. So the best-fit alignment becomes closely similar to that of SNe 1986, 2006X and 2008fp (model 2), which indicates a {\it universal} alignment of interstellar grains along the lines of sight to these SNe Ia. Previous studies using only extinction fitting (\cite{2015ApJ...807L..26G}; \citealt{2016arXiv160806689N}) also found the excess of small grains toward these SNe.

{Lastly, the remaining question is that why do the environments toward these SNe Ia have enhanced relative abundance of small grains?

High relative abundance of small grains may arise from several processes, including galaxy evolution, dust formation in SN II and asymptotic giant branch (AGB) stars, and dust evolution (see \citealt{Patat:2015bb}). Here we suggest that small grains may be replenished due to cloud-cloud collisions induced by strong radiation pressure of SNe Ia. 

Indeed, grains in a molecular cloud initially located closer to the explosion site are accelerated by strong SN radiation pressure to high terminal velocities, as given by $v\simeq 171 (L_{\rm bol}/10^{8}L_{\odot})^{1/2}(r_{i}/100\pc)^{-1/2}a_{-5}^{-1/2} \km\s^{-1}$ (see \citealt{Hoang:2015bn}), and this cloud may be ejected as well. The ejecting cloud would collide with another cloud in its vicinity (e.g., for SN 1986G and 2014J), which certainly results in destruction of big grains. It is noted that the drifting motion of fast grains through the ambient gas results in non-thermal sputtering of the grains, which reduces the abundance of small grains. But this process usually takes a long timescale to be important, i.e., $\tau_{\rm sp}\sim a\rho/n_{\gas}v_{\rm gr} m_{\H}\sim 500 a_{-5}\rho/3\g\cm^{-3}/(n_{\gas}/100\cm^{-3})(v_{\rm gr}/100\km\s^{-1})$ yr. {Therefore, these scenarios have little impact on the time variability of extinction and polarization toward SNe that are usually carried out within several weeks after the explosions.}

\subsection{Alignment of interstellar dust grains toward the SNe Ia}
\subsubsection{SNe Ia with low values of $\lambda_{\max}\sim 0.3-0.45\mum$}

The polarization curves for SNe 1986G and 2006X exhibit low values of $\lambda_{\max}\sim 0.3-0.45\mum$. For SNe 2008fp and 2014J, when accounting for the effect of Rayleigh scattering (model 2), we found that their values of $\lambda_{\max}$ become less extreme, as small as in SNe 1986G and 2006X. To reproduce such low values of $\lambda_{\max}$, we find that the enhancement in relative abundance of small silicate grains is sufficient, while the alignment function exhibits a normal behavior in which grains larger than $a\sim 0.05\mum$ are efficiently aligned as in the Galaxy (see Figure \ref{fig:nafali1}). This normal alignment of interstellar grains is most likely possible via radiative torque (RAT) mechanism induced by interstellar radiation fields (see \citealt{LAH15} for a review). It is noted that both SNe 1986G and 2014J exploded in the disk of starburst galaxies NGC 5128 and NGC 3034, and there exist numerous clouds along the LOS. Therefore, the grain alignment in these cases would reflect the alignment properties of dust grains in the average ISM of the host galaxies.

\subsubsection{SNe Ia with extreme values of $\lambda_{\max}<0.2\mum$: radiative torque alignment by radiation from SNe Ia}

Among the SNe Ia considered, SN 2008fp is an exceptional case in which both model 1 and model 2 could reproduce the observational data. Model 1 without CS dust provides a slightly better fit (smaller $\chi^{2}$), but its extreme value of $\lambda_{\max}=0.15\mum$ requires that small grains ($a\sim 0.01$) are aligned as efficient as big grains. This is unusual because small grains are usually found to be weakly aligned, e.g., in the Galaxy (\citealt{1995ApJ...444..293K}; HLM14). Therefore, it is important to understand whether such a special grain alignment is physically possible.

Over the past few years, we have witnessed significant progress in theoretical and observational studies of grain alignment, and RAT alignment has become a leading mechanism for the alignment of interstellar grains  (see \citealt{Andersson:2015bq} and \citealt{LAH15} for reviews). According to the RAT model, the alignment of irregular grains depends on a number of physical parameters, including the grain size, , the radiation energy density, mean wavelength of the radiation spectrum, the angle between the anisotropic direction of the radiation field and the magnetic field, and the local gas density and temperature (\citealt{2007MNRAS.378..910L}; \citealt{{Hoang:2008gb},{2009ApJ...697.1316H}}). For a SN of luminosity $L_{\rm SN}$, the radiation energy density at distance $d_{pc}$ in units of pc is given by
\bea
u_{\rm rad}&=&\int u_{\lambda}d\lambda =\int  \frac{L_{\lambda}e^{-\tau_{\lambda}}}{4\pi c d^{2}}d\lambda,\nonumber\\
&\simeq& 1.06\times 10^{-7}\frac{L_{8}e^{-\tau}}{d_{pc}^{2}} \erg \cm^{-3},\label{eq:urad}
\ena
where $u_{\lambda}$ is the spectral energy density, $\tau_{\lambda}$ is the optical depth induced by intervening dust, $L_{8}=L_{\rm SN}/10^{8}L_{\odot}$, and $\tau$ is defined as $e^{-\tau}=\int L_{\lambda}e^{-\tau_{\lambda}}d\lambda/L_{\rm SN}$. 

Thus, the critical size of above which grains are efficiently aligned, $a_{\rm ali}$, is determined by the maximum angular momentum induced by RATs, which is equal to (see \citealt{2007MNRAS.378..910L}; \citealt{2014MNRAS.438..680H}):
\bea
\frac{J_{\rm max}}{J_{\rm th}}&\approx &30\hat{\gamma}_{\rm rad}\hat{\rho}^{1/2}a_{-5}^{1/2}
\left(\frac{10^{3}\rm cm^{-3}}{n_{\rm H}}\right)\left(\frac{20\rm K}{T_{\rm gas}}\right)\nonumber\\
&&\times
\left(\frac{\bar{\lambda}}
{1.2\mum}\right)\left(\frac{10^{6}L_{8}e^{-\tau}}{d_{pc}^{2}}\right)\left(\frac{\overline{Q_{\Gamma}}}{10^{-2}}\right)
\left(\frac{1}{1+F_{\rm IR}}\right),~~~~~\label{eq:Jmax_RAT}
\ena
where $J_{\rm th}$ is the thermal angular momentum of grains at gas temperature, $F_{\rm IR}$ is the damping coefficient due to infrared emission, $\hat{\gamma}_{\rm rad}=\gamma_{\rm rad}/0.1$ with $\gamma_{\rm rad}$ the anisotropy degree of the radiation field (unity in our case), $
\bar{\lambda}$ and $\overline{Q}_{\Gamma}$ are the wavelength and RAT efficiency averaged over the entire radiation field spectrum, respectively. Here we disregard the effect of the ISRF, which is small compared to the SN radiation in the clouds close to the SN.

For interstellar grains with $a\ll \overline{\lambda}$, $\overline{Q}_{\Gamma}$ is approximately equal to (\citealt{2014MNRAS.438..680H})
\bea
\overline{Q}_{\Gamma}\simeq 2\left(\frac{\overline{\lambda}}{a}\right)^{-2.7}\simeq 2.4\times 10^{-3}\left(\frac{\overline{\lambda}}{1.2\mum}\right)^{-2.7}a_{-5}^{2.7}.
\ena

Grains are expected to be efficiently aligned when spun up to suprathermal speeds, which is determined by the criteria $J_{\max}\ge 3J_{\rm th}$ (see \citealt{Hoang:2008gb}). The timescale that RATs can spin-up grains to suprathermal rotation is equal to
\bea
\tau_{\rm spin-up}&\equiv& \frac{J_{\rm th}}{dJ/dt}= \frac{\tau_{\rm drag}}{J_{\max}/J_{\rm th}},\nonumber\\
&\simeq& \frac{6.58\times 10^{4}a_{-5}n_{1}^{-1}T_{2}^{1/2}yr}{J_{\max}/J_{\rm th}},
\ena
where $\tau_{\rm drag}\simeq 6.58\times 10^{4}a_{-5}n_{1}^{-1}T_{2}^{1/2}/(1+F_{\rm IR})$yr with $n_{1}=n_{\H}/10\cm^{-3}$ is the rotational damping time due to gas drag. Plugging numerical values, we get
\bea
\tau_{\rm spin-up}\simeq 0.5 \left(\frac{d_{pc}^{2}}{L_{8}e^{-\tau}} \right)\bar{\lambda}^{1.7}a_{-5}^{-2.2} {~\rm day}.
\ena

{Radiation energy of SNe Ia is produced mostly by conversion of the kinetic energy of ejecta interacting with surrounding environments (i.e., shocked regions). Most of such energy is concentrated in UV-optical wavelengths, especially in early epochs after the explosion (see \citealt{2009AJ....137.4517B}). For simplicity, let assume that the emitting region can be approximately described by a black body of effective temperature $T_{\rm SN}$. Thus, we can estimate $\bar{\lambda}=\int \lambda u_{\lambda}(T_{\rm SN}) d\lambda/\int u_{\lambda}(T_{\rm SN})d\lambda$. This gives $\bar{\lambda}=0.35\mum$ for $T_{\rm SN}=1.5\times10^{4}$K.}


\begin{figure*}
\centering
\includegraphics[width=0.4\textwidth]{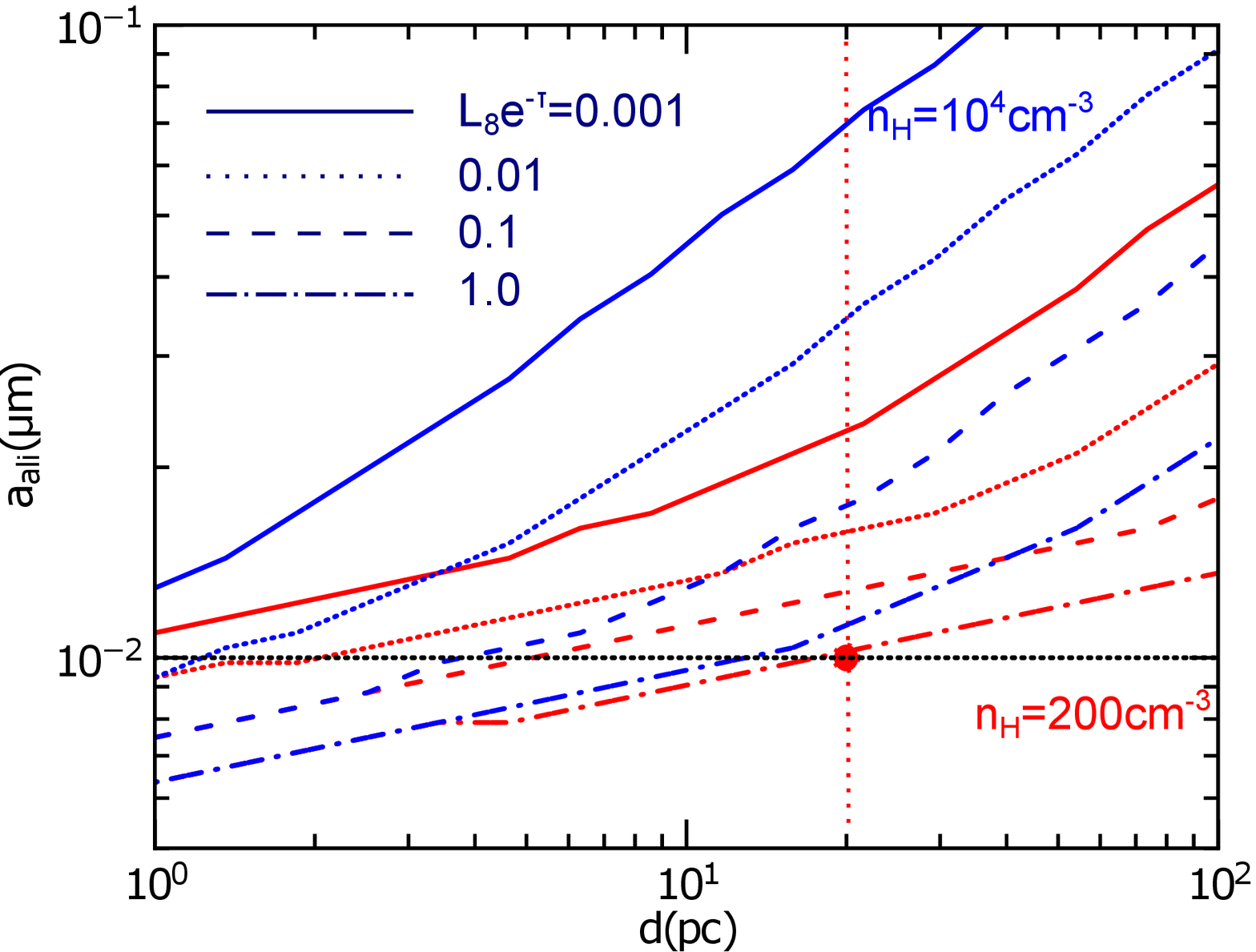}
\includegraphics[width=0.4\textwidth]{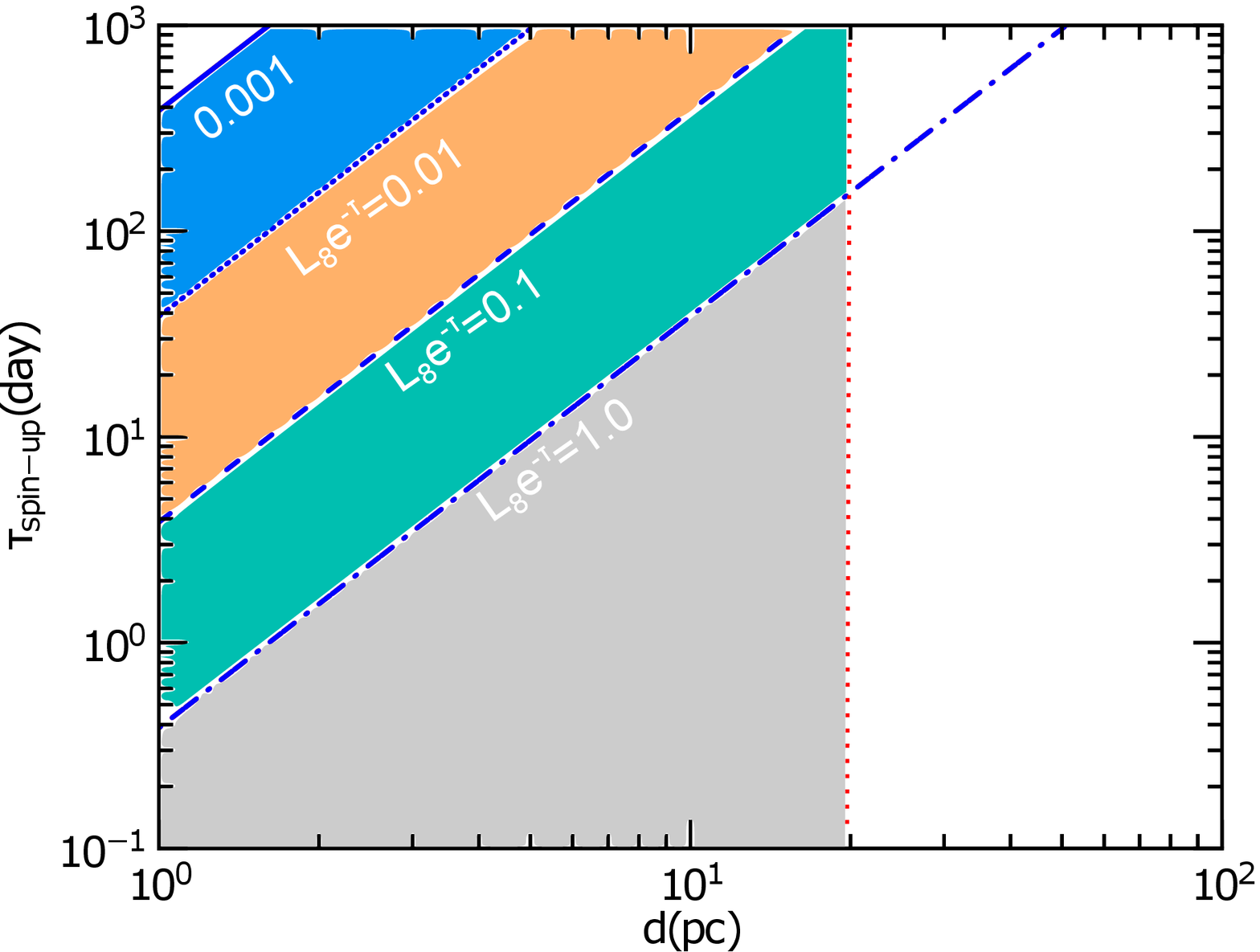}
\caption{Left panel: critical size of aligned grains $a_{\rm ali}$ by RATs as a function of the cloud distance $d$ for different values of $L_{8}e^{-\tau}=10^{-3}-1$. Two values of gas density $n_{\H}=200\cm^{-3}$ (red) and $10^{4}\cm^{-3}$ (blue) are considered. Dotted vertical lines indicates the alignment $a_{\ali}=0.01\mum$ for $L_{8}e^{-\tau}=1$ and $n_{H}=200\cm^{-3}$. Right panel: spin-up time $\tau_{spin-up}$ for $a=0.01\mum$. Shaded areas mark the region where the RAT alignment of $a\ge 0.01\mum$ grains can be observed for a given pair $L_{8}e^{-\tau}$ and $d_{pc}$.}
\label{fig:aali}
\end{figure*}

In Figure \ref{fig:aali} we shows $a_{\ali}$ (left panel) and $\tau_{\rm spin-up}$ (right panel) as functions of $d_{\rm pc}$ for different values of $n_{\H}$ and $L_{8}e^{-\tau}$. The left panel shows that small grains $a_{\rm ali}\sim 0.01\mum$ in the cloud at distances $d \sim 20$ pc with the gas density $n_{\H}\sim 250\cm^{-3}$ (as estimated for SN 2008fp in \citealt{Cox:2014fq}) could be aligned. For a very dense cloud $n_{\H}\sim 10^{4}\cm^{-3}$, it must be very close at $d\sim 3$ pc to give $a_{\ali}\sim 0.02\mum$ for $L_{8}e^{-\tau}<1$. These small distances compared to the galaxy scale indicate that the clouds are close to the explosion site to reproduce the observed polarization. 

One important feature of the RAT alignment is the spin-up time, which may be longer than the observation time since the explosion time. From Figure \ref{fig:aali} it shows that for a cloud at distance $d\sim 1-10pc$ (i.e., well beyond the sublimation radius, see \citealt{Hoang:2015bn}), the $a\sim 0.03\mum$ grains can be radiatively aligned on $t_{\rm spin-up}\sim 0.3-30$ days for $L_{\rm SN}=10^{8}L_{\odot}$. 

It is noted that polarimetric observations of SN 2008fp were performed on -2, 3, 9 and 31 days relative to the maximum epoch \citep{Cox:2014fq}.  As a result, RAT alignment by the radiation from SNe requires the cloud be within 10 pc such that $t_{\rm obs}>\tau_{\rm spin-up}$. {For SN 2014J, the observations were carried out on three epochs (Jan 28, Feb 3, Mar 8, 2014) with the maximum brightness in the first week of February (\citealt{Patat:2015bb}). The situation is more complicated because the LOS toward this SN contains more than 20 clouds.}

One related issue is that if the RAT alignment by SNe Ia's radiation is at work, then why do other SNe Ia not exhibit extreme values of $\lambda_{\max}$? The first reason may be that this cloud is located further away than the single cloud in SN 2008fp, or the radiation from the central source is significantly reduced by larger optical depth $\tau$ arising from circumstellar dust (e.g. for SN 2014J) or intervening clouds. 

{Lastly, our discussion of RAT alignment of grains in intervening clouds above can be applied to CS dust; the only difference is that in the latter the radiation from white dwarf (WD) induces grain alignment. {Our estimate for a typical WD with $T_{\star}\sim 10^{5}$K and $R_{\star}\sim 0.01R_{\odot}$ shows that grains can be aligned within 10R$_{\star}$. Nevertheless, those grains will be swept out rapidly after SN explosion (see Section \ref{sec:model}) and have no impact on the polarization degree and dispersion angles through the SN explosion epochs.}

\subsubsection{Observational signatures of RAT alignment by SNe Ia's radiation}
The variability of the SN luminosity results in the variation of $a_{\ali}$, which apparently induces the variation of the polarization curves. To illustrate such an effect, we compute a number of polarization curves by varying $a_{\rm ali}$, where a typical grain size distribution of the Galaxy (\citealt{2007ApJ...657..810D}) is chosen. The alignment function is assumed to be $f_{a}=1-\exp[-\left(a/a_{\rm ali}\right)^{3}]$, which captures our best-fit alignment functions shown in Figure \ref{fig:nafali1}. Obtained results are shown in Figure \ref{fig:polSN}. It is shown that, when $a_{\ali}$ is decreased from $0.03\mum$ to $0.001\mum$, the polarization at $\lambda<0.4\mum$ varies substantially, whereas its variation at $\lambda>0.4\mum$ is rather small.

The increase and decrease of the polarization at $\lambda<0.4\mum$ before and after the maximum luminosity epoch is a unique signature of RAT alignment toward SNe. Therefore, observing the far-UV SN polarization would provide important insight into the origin of anomalous polarization and the RAT alignment mechanism.  Because the IS polarization is determined by the interstellar magnetic fields, the polarization angles are stable through the different epochs of SN observations. Thus, this model of RAT alignment naturally explains the lack of the variability of polarization vectors (see \citealt{Patat:2015bb}; \citealt{2016ApJ...828...24P}).

\begin{figure}
\includegraphics[width=0.4\textwidth]{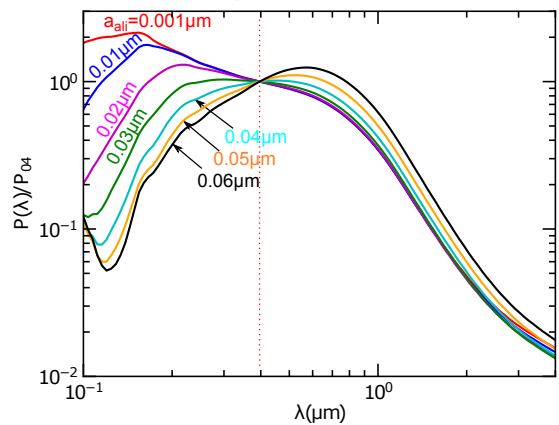}
\caption{Normalized polarization curves predicted for different values of $a_{\ali}$: peak wavelength shifts to blue when $a_{\ali}$ is decreased. The contribution of small grains of $a\le 0.03\mum$ is negligible for the polarization at $\lambda\ge 0.4\mum$.}
\label{fig:polSN}
\end{figure}

\subsection{Circumstellar dust and implications for SNe Ia progenitors}\label{sec:model}

Using inversion technique, we found that the presence of CS dust in SN 2008fp is uncertain because our inversion is successful for both models without and with CS dust. Including CS dust allows us to get dust parameters $R_{V}, \lambda_{\max}$ to be similar for four SNe Ia, and the alignment function of the IS dust toward these SNe Ia appears to be universal. However, it has been argued that CS dust is unlikely important for SN 2008fp due to the lack of variability in the polarization angles (\citealt{Patat:2015bb}; \citealt{2016ApJ...816...57M}) as well as the lack of variability of extragalactic absorption line profiles (see \citealt{Cox:2014fq}).

The existence of CS dust in SN2014J is still strongly debated. Some groups claim that the extinction to SN 2014J is dominated by interstellar dust (\citealt{Patat:2015bb}; \citealt{2015ApJ...805...74B}), whereas others (\citealt{Foley:2014br}) suggest the CS dust may be important. Our model with CS dust can be successfully obtained, indicating that the CS dust is likely important. This is in agreement with a recent analysis of M82 region before SN 2014J by \cite{2015arXiv150603821H} that supports the contribution of both CS dust and IS dust for low $R_{V}$. 

\cite{Patat:2015bb} argued that the presence of CS dust in SN 2014J may not be ruled out, although the lack of variability in the polarization degree and polarization angles is difficult to reconcile. {The authors suggested that if scattering is a dominant source of observed polarization, then the geometry of CS dust cloud has to be such that the resulting polarization angle, by coincidence, is aligned with that of the local magnetic field.}

{A possible source that can explain the invariability of the polarization is a WD containing an accretion disk within its magnetosphere (private communication with B-G Andersson). Small dust grains in the disk then undergo Rayleigh scattering with the polarization perpendicular to the disk plane. If the accretion disk is parallel to the Galactic plane, the polarization by scattering is parallel to the interstellar polarization, successfully explaining the invariability of the polarization angles. However, dust grains in the accretion disk and magnetosphere would likely be swept away by SN ejecta shortly after the explosion, which is a serious challenge for this toy model.}

Indeed, the WD magnetosphere is characterized by Alfven radius at which the magnetic energy is balanced with the gas thermal energy, which is given by:
\bea
 R_{A}&\sim& \left(3B_{\star}^{2}R_{\star}^{6}/2\dot{M}\sqrt{GM_{\star}}\right)^{2/7},\nonumber\\
 &\simeq& 1.2R_{\odot}B_{\star}(kG)^{4/7}(\dot{M}/10^{8}\g\s^{-1})^{-2/7}\nonumber\\
 &&\times(R_{\star}/0.01R_{\odot})^{12/7}(M_{\star}/0.6M_{\odot})^{-1/7},
 \ena
 where $\dot{M}$ is the accretion rate, and $B_{\star}$ is the magnetic field (see \citealt{2012MNRAS.423..505M}). For typical parameters $\dot{M}\sim 10^{6}-10^{10}\g\s^{-1}$ and $B_{\star}\sim 1-100$ kG, it yields $R_{A}\sim 0.3-62R_{\odot}\times(R_{\star}/0.01R_{\odot})^{12/7}(M_{\star}/0.6M_{\odot})^{-1/7}$.
 
It can be seen that dust grains within the magnetosphere would be swept away after $t\simeq 100 R_{\odot}/v_{\rm ej} \sim 2.1 (v_{\rm ej}/0.03c)^{-1}$ hr, where the ejecta velocity during the free-expansion period is $v_{\rm ej}\sim 10^{4}\km\s^{-1}$. As a result, observations at hours after the SN explosion would not see any signatures of such dust.


\subsection{Magnetic alignment and Implication for magnetic fields in host galaxies}
It is worth to mention an alternative paramagnetic alignment mechanism that is proposed by \cite{1951ApJ...114..206D} to explain starlight polarization. However, paramagnetic alignment of thermally rotating small grains is found to be inefficient (\citealt{1997MNRAS.288..609L}; \citealt{2014ApJ...790....6H}). Without suprathermal rotation, even grains with superparamagnetic inclusions are far from being efficiently aligned (\citealt{1999MNRAS.305..615R}; Hoang \& Lazarian 2015, submitted). The joint action of RATs and superparamagnetic inclusions would result in the perfect alignment of sufficiently large grains \citep{Lazarian:2008fw}.

\cite{2014ApJ...790....6H} found that increasing the strength of magnetic fields can produce enhancement of alignment of small grains, which partly results in the low values of $\lambda_{\max}$ and account for the polarization excess in far-UV ($\lambda<0.2\mum$). Considering the environment conditions of four SNe Ia under interest, we see that SNe 1986G and 2014J belong to the starburst galaxies, which usually exhibit stronger magnetic fields than normal galaxies \citep{2006ApJ...645..186T}. Thus, the low values of $\lambda_{\max}$ perhaps arise from the high magnetic fields as predicted by HLM14. If this is the case, far-UV polarimetric observations using SNe Ia would provide valuable constraints on the interstellar magnetic fields of external galaxies.

The interstellar magnetic field along the lines of sight toward SNe Ia likely undergoes wandering due to interstellar turbulence. In our inversion technique, the effect of the magnetic field wandering is incorporated into the alignment function $f=\Phi R\cos^{2}\beta$, where $\beta$ is the angle between the mean regular field and the plane of the sky, and $\Phi$ describes the average fluctuations of the local field with the LOS. The maximum value of $f$ is constrained such that $P_{\max}/A(\lambda_{\max})=3\%/mag$ corresponds to $f_{\max}=1$.

{Lastly, for our modeling, we have not fitted to real observational data but to the analytical fits (CCM and Serkowski laws) of the observations. The reason for that is that we have attempted to infer dust properties and grain alignment for the wide range of grain size, from ultrasmall to large grains. Due to the lack of UV observations, the analytical fits are useful to extrapolate for the UV data from the available data in optical and near-IR. As a result, the caveat of our study is the uncertainty in the inferred values $R_{V}$ and $\lambda_{\max}$ due to the lack of UV data. However, our modeling will be updated easily when future UV observations are available.}

\section{Summary}\label{sec:summ}
In the present paper, we have studied the properties and alignment of interstellar dust grains in external galaxies through four SNe Ia. The main findings are summarized as follows:

\begin{itemize}
\item[1.] Using inversion technique, we have obtained the best-fit grain size distribution and alignment function that simultaneously reproduce the observed extinction and polarization of four SNe Ia (1986G, 2006X, 2008fp, and 2014J) with the anomalous values of $R_{V}$ and $\lambda_{\max}$.

\item[2.] The best-fit grain size distributions reveal an enhancement in the mass of small silicate grains with peak around $a\sim 0.06\mum$. It is suggested that the enhanced relative abundance of small grains may be produced by cloud-cloud collisions induced by SN radiation pressure.

\item[3.] For model 1 of SN 2008fp (without CS dust), we find that to reproduce the observational data, increasing the relative abundance of small grains is insufficient. Instead, the alignment of small grains ($a\sim 0.01\mum$) must be as efficient as big grains. {Including the effect of CS dust, the alignment function becomes more reasonable, similar to that for SNe 1986G and 2006X. The fit is however not improved, and the existence of CS dust is still uncertain.}

\item[4.] We have suggested a model of grain alignment based on radiative torques induced by direct, strong radiation from SNe, which can adequately explain the efficient alignment of small grains for model 1 of SN 2008fp. Far-UV polarimetric observations would be useful to differentiate the effect of IS dust and CS dust based on its polarization spectra.

\item[5.] The existence of CS dust around SN 2014J is suggested by our successful inversion when both CS dust and IS dust must be accounted for the observed extinction and polarization.

\end{itemize}

\section*{Acknowledgments}
We thank the anonymous referee for a careful reading and helpful comments that helped us improve the paper significantly. We are grateful to B-G Andersson for insightful comments on the manuscript, and for making the suggestion that the steep rise in the observed SNe Ia polarization spectra might be due to Rayleigh scattering in the circumstellar environment. We thank Peter G Martin for interesting discussions and Nguyen-Luong Quang for useful comments and suggestions. Part of this work was done during our stay as Alexander von Humboldt Fellow at Ruhr Universit$\ddot{\rm a}$t Bochum and Goethe Universit$\ddot{\rm a}$t Frankfurt am Main.

\bibliography{ms.bbl}

\begin{thebibliography}{65}
\expandafter\ifx\csname natexlab\endcsname\relax\def\natexlab#1{#1}\fi

\bibitem[{Amanullah {et~al.}(2015)Amanullah, Johansson, Goobar, \&
  et~al.}]{2015MNRAS.453.3300A}
Amanullah, R., Johansson, J., Goobar, A., \& et~al. 2015, \mnras, 453, 3300

\bibitem[{Anders \& Grevesse(1989)}]{1989GeCoA..53..197A}
Anders, E., \& Grevesse, N. 1989, Geochimica et Cosmochimica Acta, 53, 197

\bibitem[{Anderson {et~al.}(1996)Anderson, Weitenbeck, Code, Nordsieck, Meade,
  Babler, Zellner, Bjorkman, Fox, Johnson, Sanders, Lupie, \&
  Edgar}]{1996AJ....112.2726A}
Anderson, C.~M., Weitenbeck, A.~J., Code, A.~D., {et~al.} 1996, \aj, 112, 2726

\bibitem[{Andersson {et~al.}(2015)Andersson, Lazarian, \&
  Vaillancourt}]{Andersson:2015bq}
Andersson, B.-G., Lazarian, A., \& Vaillancourt, J.~E. 2015, Annual Review of
  Astronomy and Astrophysics, 53, 501

\bibitem[{Andersson \& Potter(2007)}]{2007ApJ...665..369A}
Andersson, B.-G., \& Potter, S.~B. 2007, \apj, 665, 369

\bibitem[{Andersson {et~al.}(2013)Andersson, Piirola, De~Buizer, Clemens,
  Uomoto, Charcos-Llorens, Geballe, Lazarian, Hoang, \&
  Vornanen}]{2013ApJ...775...84A}
Andersson, B.-G., Piirola, V., De~Buizer, J., {et~al.} 2013, \apj, 775, 84

\bibitem[{Brown {et~al.}(2009)Brown, Holland, Immler, Milne, Roming, Gehrels,
  Nousek, Panagia, Still, \& Vanden~Berk}]{2009AJ....137.4517B}
Brown, P.~J., Holland, S.~T., Immler, S., {et~al.} 2009, \aj, 137, 4517

\bibitem[{Brown {et~al.}(2014)Brown, Smitka, Wang, Breeveld, de~Pasquale,
  Hartmann, Krisciunas, Kuin, Milne, Page, \& Siegel}]{2014arXiv1408.2381B}
Brown, P.~J., Smitka, M.~T., Wang, L., {et~al.} 2014, arXiv.org, 2381

\bibitem[{Brown {et~al.}(2015)Brown, Smitka, Wang, Breeveld, de~Pasquale,
  Hartmann, Krisciunas, Kuin, Milne, Page, \& Siegel}]{2015ApJ...805...74B}
Brown, P.~J., Smitka, M.~T., Wang, L., {et~al.} 2015, \apj, 805, 74

\bibitem[{Cardelli {et~al.}(1989)Cardelli, Clayton, \&
  Mathis}]{1989ApJ...345..245C}
Cardelli, J.~A., Clayton, G.~C., \& Mathis, J.~S. 1989, \apj, 345, 245

\bibitem[{Chiar {et~al.}(2006)Chiar, Adamson, Whittet, Chrysostomou, Hough,
  Kerr, Mason, Roche, \& Wright}]{2006ApJ...651..268C}
Chiar, J.~E., Adamson, A.~J., Whittet, D. C.~B., {et~al.} 2006, \apj, 651, 268

\bibitem[{Clayton {et~al.}(2003)Clayton, Wolff, Sofia, Gordon, \&
  Misselt}]{2003ApJ...588..871C}
Clayton, G.~C., Wolff, M.~J., Sofia, U.~J., Gordon, K.~D., \& Misselt, K.~A.
  2003, \apj, 588, 871

\bibitem[{Cox \& Patat(2008)}]{2008A&A...485L...9C}
Cox, N. L.~J., \& Patat, F. 2008, A\&A, 485, L9

\bibitem[{Cox \& Patat(2014)}]{Cox:2014fq}
Cox, N. L.~J., \& Patat, F. 2014, A\&A, 565, A61

\bibitem[{Cristiani {et~al.}(1992)Cristiani, Cappellaro, Turatto, Bergeron,
  Bues, Buson, Danziger, di~Serego~Alighieri, Duerbeck, Heydari-Malayeri,
  Krautter, Schmutz, \& Schulte-Ladbeck}]{1992A&A...259...63C}
Cristiani, S., Cappellaro, E., Turatto, M., {et~al.} 1992, A\&A, 259, 63

\bibitem[{Davis \& Greenstein(1951)}]{1951ApJ...114..206D}
Davis, L.~J., \& Greenstein, J.~L. 1951, \apj, 114, 206

\bibitem[{D'Odorico {et~al.}(1989)D'Odorico, di~Serego~Alighieri, Pettini,
  Magain, Nissen, \& Panagia}]{1989A&A...215...21D}
D'Odorico, S., di~Serego~Alighieri, S., Pettini, M., {et~al.} 1989, A\&A, 215,
  21

\bibitem[{Draine(2003)}]{2003ARA&A..41..241D}
Draine, B.~T. 2003, \araa, 41, 241

\bibitem[{Draine \& Allaf-Akbari(2006)}]{2006ApJ...652.1318D}
Draine, B.~T., \& Allaf-Akbari, K. 2006, \apj, 652, 1318

\bibitem[{Draine \& Fraisse(2009)}]{Draine:2009p3780}
Draine, B.~T., \& Fraisse, A.~A. 2009, \apj, 696, 1

\bibitem[{Draine \& Li(2007)}]{2007ApJ...657..810D}
Draine, B.~T., \& Li, A. 2007, \apj, 657, 810

\bibitem[{Draine {et~al.}(2007)Draine, Dale, Bendo, Gordon, Smith, Armus,
  Engelbracht, Helou, Kennicutt, Li, Roussel, Walter, Calzetti, Moustakas,
  Murphy, Rieke, Bot, Hollenbach, Sheth, \& Teplitz}]{2007ApJ...663..866D}
Draine, B.~T., Dale, D.~A., Bendo, G., {et~al.} 2007, \apj, 663, 866

\bibitem[{Foley {et~al.}(2014)Foley, Fox, McCully, Phillips, Sand, Zheng,
  Challis, Filippenko, Folatelli, Hillebrandt, Hsiao, Jha, Kirshner, Kromer,
  Marion, Nelson, Pakmor, Pignata, Ropke, Seitenzahl, Silverman, Skrutskie, \&
  Stritzinger}]{Foley:2014br}
Foley, R.~J., Fox, O.~D., McCully, C., {et~al.} 2014, \mnras, 443, 2887

\bibitem[{{Fossey} {et~al.}(2014){Fossey}, {Cooke}, {Pollack}, {Wilde}, \&
  {Wright}}]{2014CBET.3792....1F}
{Fossey}, S.~J., {Cooke}, B., {Pollack}, G., {Wilde}, M., \& {Wright}, T. 2014,
  CBET, 3792, 1

\bibitem[{Gao {et~al.}(2015)Gao, Jiang, Li, Li, \& Wang}]{2015ApJ...807L..26G}
Gao, J., Jiang, B.~W., Li, A., Li, J., \& Wang, X. 2015, \apjl, 807, L26

\bibitem[{Goobar(2008)}]{2008ApJ...686L.103G}
Goobar, A. 2008, \apj, 686, L103

\bibitem[{Hoang \& Lazarian(2008)}]{Hoang:2008gb}
Hoang, T., \& Lazarian, A. 2008, \mnras, 388, 117

\bibitem[{Hoang \& Lazarian(2009)}]{2009ApJ...697.1316H}
Hoang, T., \& Lazarian, A. 2009, \apj, 697, 1316

\bibitem[{Hoang \& Lazarian(2014)}]{2014MNRAS.438..680H}
Hoang, T., \& Lazarian, A. 2014, \mnras, 438, 680

\bibitem[{Hoang {et~al.}(2013)Hoang, Lazarian, \& Martin}]{2013ApJ...779..152H}
Hoang, T., Lazarian, A., \& Martin, P.~G. 2013, \apj, 779, 152

\bibitem[{Hoang {et~al.}(2014)Hoang, Lazarian, \& Martin}]{2014ApJ...790....6H}
Hoang, T., Lazarian, A., \& Martin, P.~G. 2014, \apj, 790, 6

\bibitem[{Hoang {et~al.}(2015)Hoang, Lazarian, \& Schlickeiser}]{Hoang:2015bn}
Hoang, T., Lazarian, A., \& Schlickeiser, R. 2015, \apj, 806, 1

\bibitem[{Hough {et~al.}(1987)Hough, Bailey, Rouse, \&
  Whittet}]{Hough:1987p6065}
Hough, J.~H., Bailey, J.~A., Rouse, M.~F., \& Whittet, D. C.~B. 1987, \mnras,
  227, 1P

\bibitem[{Hutton {et~al.}(2015)Hutton, Ferreras, \&
  Yershov}]{2015arXiv150603821H}
Hutton, S., Ferreras, I., \& Yershov, V. 2015, arXiv:1506.03821, 3821

\bibitem[{Kawabata {et~al.}(2014)Kawabata, Akitaya, Yamanaka, Itoh, Maeda,
  Moritani, Ui, Kawabata, Mori, Nogami, Nomoto, Suzuki, Takaki, Tanaka, Ueno,
  Chiyonobu, Harao, Matsui, Miyamoto, Nagae, Nakashima, Nakaya, Ohashi, Ohsugi,
  Komatsu, Sakimoto, Sasada, Sato, Tanaka, Urano, Yamashita, Yoshida, Arai,
  Ebisuda, Fukazawa, Fukui, Hashimoto, Honda, Izumiura, Kanda, Kawaguchi,
  Kawai, Kuroda, Masumoto, Matsumoto, Nakaoka, Takata, Uemura, \&
  Yanagisawa}]{Kawabata:2014gy}
Kawabata, K.~S., Akitaya, H., Yamanaka, M., {et~al.} 2014, \apj, 795, L4

\bibitem[{Kim \& Martin(1995)}]{1995ApJ...444..293K}
Kim, S.-H., \& Martin, P.~G. 1995, \apj, 444, 293

\bibitem[{Kim {et~al.}(1994)Kim, Martin, \& Hendry}]{1994ApJ...422..164K}
Kim, S.-H., Martin, P.~G., \& Hendry, P.~D. 1994, \apj, 422, 164

\bibitem[{Larson {et~al.}(2000)Larson, Larson, Wolff, Wolff, Roberge, Roberge,
  Whittet, Whittet, He, \& He}]{2000ApJ...532.1021L}
Larson, K.~A., Larson, K.~A., Wolff, M.~J., {et~al.} 2000, \apj, 532, 1021

\bibitem[{Lazarian(1997)}]{1997MNRAS.288..609L}
Lazarian, A. 1997, \mnras, 288, 609

\bibitem[{{Lazarian} {et~al.}(2015){Lazarian}, {Andersson}, \& {Hoang}}]{LAH15}
{Lazarian}, A., {Andersson}, B.-G., \& {Hoang}, T. 2015, in Polarimetry of
  stars and planetary systems, ed. L.~{Kolokolova}, J.~{Hough}, \& A.-C.
  {Levasseur-Regourd} (New York:Cambridge University Press), 81

\bibitem[{Lazarian \& Hoang(2007)}]{2007MNRAS.378..910L}
Lazarian, A., \& Hoang, T. 2007, \mnras, 378, 910

\bibitem[{Lazarian \& Hoang(2008)}]{Lazarian:2008fw}
Lazarian, A., \& Hoang, T. 2008, \apj, 676, L25

\bibitem[{Maeda {et~al.}(2016)Maeda, Tajitsu, Kawabata, Foley, Honda, Moritani,
  Tanaka, Hashimoto, Ishigaki, Simon, Phillips, Yamanaka, Nogami, Arai, Aoki,
  Nomoto, Milisavljevic, Mazzali, Soderberg, Schramm, Sato, Harakawa, Morrell,
  \& Arimoto}]{2016ApJ...816...57M}
Maeda, K., Tajitsu, A., Kawabata, K.~S., {et~al.} 2016, \apj, 816, 57

\bibitem[{Metzger {et~al.}(2012)Metzger, Rafikov, \&
  Bochkarev}]{2012MNRAS.423..505M}
Metzger, B.~D., Rafikov, R.~R., \& Bochkarev, K.~V. 2012, \mnras, 423, 505

\bibitem[{Nobili \& Goobar(2008)}]{2008A&A...487...19N}
Nobili, S., \& Goobar, A. 2008, A\&A, 487, 19

\bibitem[{Nozawa(2016)}]{2016arXiv160806689N}
Nozawa, T. 2016, arXiv.org, arXiv:1608.06689

\bibitem[{Parkin {et~al.}(2012)Parkin, Wilson, Foyle, Baes, Bendo, Boselli,
  Boquien, Cooray, Cormier, Davies, Eales, Galametz, Gomez, Lebouteiller,
  Madden, Mentuch, Page, Pohlen, Remy, Roussel, Sauvage, Smith, \&
  Spinoglio}]{2012MNRAS.422.2291P}
Parkin, T.~J., Wilson, C.~D., Foyle, K., {et~al.} 2012, \mnras, 422, 2291

\bibitem[{Patat {et~al.}(2009)Patat, Baade, H{\"o}flich, Maund, Wang, \&
  Wheeler}]{Patat:2009fa}
Patat, F., Baade, D., H{\"o}flich, P., {et~al.} 2009, A\&A, 508, 229

\bibitem[{Patat {et~al.}(2015)Patat, Taubenberger, Cox, Baade, Clocchiatti,
  H{\"o}flich, Maund, Reilly, Spyromilio, Wang, Wheeler, \&
  Zelaya}]{Patat:2015bb}
Patat, F., Taubenberger, S., Cox, N. L.~J., {et~al.} 2015, A\&A, 577, A53

\bibitem[{Phillips {et~al.}(2013)Phillips, Simon, Morrell, Burns, Cox, Foley,
  Karakas, Patat, Sternberg, Williams, Gal-Yam, Hsiao, Leonard, Persson,
  Stritzinger, Thompson, Campillay, Contreras, Folatelli, Freedman, Hamuy,
  Roth, Shields, Suntzeff, Chomiuk, Ivans, Madore, Penprase, Perley, Pignata,
  Preston, \& Soderberg}]{2013ApJ...779...38P}
Phillips, M.~M., Simon, J.~D., Morrell, N., {et~al.} 2013, \apj, 779, 38

\bibitem[{Porter {et~al.}(2016)Porter, Leising, Williams, Milne, Smith, Smith,
  Bilinski, Hoffman, Huk, \& Leonard}]{2016ApJ...828...24P}
Porter, A.~L., Leising, M.~D., Williams, G.~G., {et~al.} 2016, \apj, 828, 24

\bibitem[{Riess {et~al.}(1998)Riess, Filippenko, Challis, \&
  et~al.}]{1998AJ....116.1009R}
Riess, A.~G., Filippenko, A.~V., Challis, P., \& et~al. 1998, \aj, 116, 1009

\bibitem[{Ritchey {et~al.}(2014)Ritchey, Welty, Dahlstrom, \&
  York}]{Ritchey:2014uq}
Ritchey, A.~M., Welty, D.~E., Dahlstrom, J.~A., \& York, D.~G. 2014, arXiv.org

\bibitem[{Roberge \& Lazarian(1999)}]{1999MNRAS.305..615R}
Roberge, W.~G., \& Lazarian, A. 1999, \mnras, 305, 615

\bibitem[{Serkowski {et~al.}(1975)Serkowski, Mathewson, \&
  Ford}]{Serkowski:1975p6681}
Serkowski, K., Mathewson, D.~S., \& Ford, V.~L. 1975, \apj, 196, 261

\bibitem[{Thompson {et~al.}(2006)Thompson, Quataert, Waxman, Murray, \&
  Martin}]{2006ApJ...645..186T}
Thompson, T.~A., Quataert, E., Waxman, E., Murray, N., \& Martin, C.~L. 2006,
  \apj, 645, 186

\bibitem[{Wang \& Wheeler(2008)}]{2008ARA&A..46..433W}
Wang, L., \& Wheeler, J.~C. 2008, Annual Review of Astronomy and Astrophysics,
  46, 433

\bibitem[{Wang {et~al.}(2008{\natexlab{a}})Wang, Li, Filippenko, Foley, Smith,
  \& Wang}]{2008ApJ...677.1060W}
Wang, X., Li, W., Filippenko, A.~V., {et~al.} 2008{\natexlab{a}}, \apj, 677,
  1060

\bibitem[{Wang {et~al.}(2012)Wang, Wang, Filippenko, \&
  et~al.}]{2012ApJ...749..126W}
Wang, X., Wang, L., Filippenko, A.~V., \& et~al. 2012, \apj, 749, 126

\bibitem[{Wang {et~al.}(2008{\natexlab{b}})Wang, Li, Filippenko, Krisciunas,
  Suntzeff, Li, Zhang, Deng, Foley, Ganeshalingam, Li, Lou, Qiu, Shang,
  Silverman, Zhang, \& Zhang}]{2008ApJ...675..626W}
Wang, X., Li, W., Filippenko, A.~V., {et~al.} 2008{\natexlab{b}}, \apj, 675,
  626

\bibitem[{Weingartner \& Draine(2001)}]{2001ApJ...548..296W}
Weingartner, J.~C., \& Draine, B.~T. 2001, \apj, 548, 296

\bibitem[{Welty {et~al.}(2014)Welty, Ritchey, Dahlstrom, \&
  York}]{2014ApJ...792..106W}
Welty, D.~E., Ritchey, A.~M., Dahlstrom, J.~A., \& York, D.~G. 2014, \apj, 792,
  106

\bibitem[{Whittet {et~al.}(1992)Whittet, Martin, Hough, Rouse, Bailey, \&
  Axon}]{Whittet:1992p6073}
Whittet, D. C.~B., Martin, P.~G., Hough, J.~H., {et~al.} 1992, \apj, 386, 562

\bibitem[{Wilking {et~al.}(1980)Wilking, Lebofsky, Kemp, Martin, \&
  Rieke}]{1980ApJ...235..905W}
Wilking, B.~A., Lebofsky, M.~J., Kemp, J.~C., Martin, P.~G., \& Rieke, G.~H.
  1980, \apj, 235, 905

\bibitem[{Zubko {et~al.}(2004)Zubko, Dwek, \& Arendt}]{2004ApJS..152..211Z}
Zubko, V., Dwek, E., \& Arendt, R.~G. 2004, \apjs, 152, 211

\end{thebibliography}
\end{document}